\documentclass[final,3p,number]{elsarticle}

\usepackage{amssymb}
\usepackage{amsmath}
\usepackage{amsthm}
\usepackage{bm}
\usepackage[colorlinks=true]{hyperref}
\usepackage[section]{placeins}
\usepackage[T1]{fontenc}
\usepackage[bitstream-charter]{mathdesign} %\usepackage{charter} with no implication on math
\usepackage[scaled]{berasans}              %sans to complement bitstream-charter
\usepackage{color}
\usepackage{graphicx}
\usepackage[normalem]{ulem}

\usepackage{esint}

\usepackage{algorithm}
\usepackage{algorithmic}
\usepackage{caption}

\usepackage{prettyref}
\newcommand\reftext[3][]{\hyperref[#2]{#3\ref{#2}#1}}
\newcommand\eqreftext[3][]{\hyperref[#2]{#3\eqref{#2}#1}}
\newrefformat{sec}{\reftext{#1}{Section \S}}
\newrefformat{ssc}{\reftext{#1}{Subsection \S}}
\newrefformat{tbl}{\reftext{#1}{Table~}}
\newrefformat{fig}{\reftext{#1}{Fig.~}}
\newrefformat{chp}{\reftext{#1}{Chapter~}}
\newrefformat{eq}{\eqreftext{#1}{Eq.~}}
\newrefformat{cp}{\eqreftext{#1}{CP~}}
\newrefformat{lcp}{\eqreftext{#1}{LCP~}}
\newrefformat{dvi}{\eqreftext{#1}{DVI~}}
\newrefformat{set}{\eqreftext{#1}{Eq.~Set~}}
\newrefformat{alg}{\reftext{#1}{Algorithm~}}
\newrefformat{rem}{\reftext{#1}{Remark~}}
\newrefformat{thm}{\reftext{#1}{Theorem~}}
\newrefformat{cor}{\reftext{#1}{Corollary~}}
\newcommand\pr[1]{\prettyref{#1}}
\newcommand{\AVE}[1]{\ensuremath{\left\langle {#1} \right\rangle}}
\newcommand{\ABS}[1]{\ensuremath{\left\lvert {#1} \right\rvert}}

\newcommand{\NORMtwo}[1]{\ensuremath{\left\lVert {#1} \right\rVert_2}}
\newcommand{\CP}[2]{\ensuremath{0\leq {#1} \perp {#2} \geq 0}}
\newcommand{\CONJ}[1]{\ensuremath{\overline{#1}}}

\newcommand{\dpone}[2]{\ensuremath{\displaystyle\frac{\partial {#1}}{\partial {#2}}}}

\newcommand{\sctf}[2]{\ensuremath{{#1}\times 10^{#2}}}
\newcommand{\Ocomp}[1]{\ensuremath{O\left( {#1} \right)}}

\newcommand{\bA}{\ensuremath{\bm{A}}}

\newcommand{\bD}{\ensuremath{\bm{D}}}

\newcommand{\bF}{\ensuremath{\bm{F}}}
\newcommand{\bG}{\ensuremath{\bm{G}}}
\newcommand{\bH}{\ensuremath{\bm{H}}}
\newcommand{\bI}{\ensuremath{\bm{I}}}

\newcommand{\bM}{\ensuremath{\bm{M}}}

\newcommand{\bT}{\ensuremath{\bm{T}}}
\newcommand{\bU}{\ensuremath{\bm{U}}}
\newcommand{\bV}{\ensuremath{\bm{V}}}
\newcommand{\bW}{\ensuremath{\bm{W}}}
\newcommand{\bX}{\ensuremath{\bm{X}}}

\newcommand{\bb}{\ensuremath{\bm{b}}}
\newcommand{\bc}{\ensuremath{\bm{c}}}

\newcommand{\be}{\ensuremath{\bm{e}}}
\newcommand{\bff}{\ensuremath{\bm{f}}}
\newcommand{\bg}{\ensuremath{\bm{g}}}

\newcommand{\bn}{\ensuremath{\bm{n}}}

\newcommand{\bp}{\ensuremath{\bm{p}}}
\newcommand{\bq}{\ensuremath{\bm{q}}}
\newcommand{\br}{\ensuremath{\bm{r}}}
\newcommand{\bs}{\ensuremath{\bm{s}}}
\newcommand{\bt}{\ensuremath{\bm{t}}}
\newcommand{\bu}{\ensuremath{\bm{u}}}
\newcommand{\bv}{\ensuremath{\bm{v}}}

\newcommand{\bx}{\ensuremath{\bm{x}}}
\newcommand{\by}{\ensuremath{\bm{y}}}

\newcommand{\bPhi}{\ensuremath{\bm{\Phi}}}
\newcommand{\bPsi}{\ensuremath{\bm{\Psi}}}
\newcommand{\bgamma}{\ensuremath{\bm{\gamma}}}

\newcommand{\bsigma}{\ensuremath{\bm{\sigma}}}

\newcommand{\bOmega}{\ensuremath{\bm{\Omega}}}
\newcommand{\btheta}{\ensuremath{\bm{\theta}}} % quaternion
\newcommand{\brho}{\ensuremath{\bm{\rho}}}

\newcommand{\bzeta}{\ensuremath{\bm{\zeta}}}
\newcommand{\btau}{\ensuremath{\bm{\tau}}}

\newcommand{\Acal}{\ensuremath{{\mathcal{A}}}}

\newcommand{\bFcal}{\ensuremath{\bm{\mathcal{F}}}}
\newcommand{\bUcal}{\ensuremath{\bm{\mathcal{U}}}}
\newcommand{\bMcal}{\ensuremath{\bm{\mathcal{M}}}}

\newcommand{\bDcal}{\ensuremath{\bm{\mathcal{D}}}}
\newcommand{\bAcal}{\ensuremath{\bm{\mathcal{A}}}}
\newcommand{\bCcal}{\ensuremath{\bm{\mathcal{C}}}}

\newcommand{\bGcal}{\ensuremath{\bm{\mathcal{G}}}}

\newcommand{\mbbR}{\mathbb{R}} % real number set

\newcommand{\Ynm}{\ensuremath{Y_n^m}}
\newcommand{\Pnm}{\ensuremath{P_n^m}}

\newcommand{\Vnm}{\ensuremath{\bV_n^m}}
\newcommand{\Wnm}{\ensuremath{\bW_n^m}}
\newcommand{\Xnm}{\ensuremath{\bX_n^m}}
\newcommand{\Gnm}{\ensuremath{\bG_n^m}}

\newcommand{\er}{\ensuremath{\be_r}}
\newcommand{\et}{\ensuremath{\be_{\theta}}}
\newcommand{\ep}{\ensuremath{\be_{\phi}}}

\newcommand{\SSg}{\mathcal{S}} % single layer
\newcommand{\SDb}{\mathcal{D}} % double layer
\newcommand{\STr}{\mathcal{K}} % traction
\newcommand{\SId}{\mathcal{I}} % identity
\newcommand{\SLo}{\mathcal{L}} % local rigid body

\begin{document}

% \verso{Wen Yan \textit{et al}}

\begin{frontmatter}
    % \title{Type the title of your paper, only capitalize first word and proper nouns\tnoteref{tnote1}}%
    \title{A scalable computational platform for particulate Stokes suspensions}%
    % \tnotetext[tnote1]{This is an example for title footnote coding.}

    \author[1]{Wen {Yan}\corref{cor1}}
    \cortext[cor1]{Corresponding author}
    \ead{wyan@flatironinstitute.org, wenyan4work@gmail.com}
    \author[3]{Eduardo {Corona}}
    \author[2]{Dhairya {Malhotra}}
    \author[4]{Shravan {Veerapaneni}}
    \author[1,2]{Michael {Shelley}}

    \address[1]{Center for Computational Biology, Flatiron Institute, Simons Foundation}
    \address[2]{Courant Institute of Mathematical Sciences, New York University}
    \address[3]{Department of Mathematics, New York Institute of Technology}
    \address[4]{Department of Mathematics, University of Michigan}

    \begin{abstract}

        We describe a computational framework for simulating suspensions of
        rigid particles in Newtonian Stokes flow.  One central building block is a
        collision-resolution algorithm that overcomes the numerical
        constraints arising from particle collisions. This algorithm extends
        the well-known complementarity method for non-smooth multi-body
        dynamics to resolve collisions in dense rigid body
        suspensions. This approach formulates the collision resolution problem
        as a linear complementarity problem with geometric `non-overlapping'
        constraints imposed at each time-step.  It is then reformulated as a
        constrained quadratic programming problem and the Barzilai-Borwein
        projected gradient descent method is applied for its solution.  This
        framework is designed to be applicable for any convex particle shape,
        e.g., spheres and spherocylinders, and applicable to any Stokes
        mobility solver, including the Rotne-Prager-Yamakawa approximation,
        Stokesian Dynamics, and PDE solvers (e.g., boundary integral and
        immersed boundary methods).  In particular, this method imposes
        Newton's Third Law and records the entire contact network.  Further,
        we describe a fast, parallel, and spectrally-accurate boundary
        integral method tailored for spherical particles, capable of resolving
        lubrication effects. We show weak and strong parallel scalings up to
        $\sctf{8}{4}$ particles with approximately $\sctf{4}{7}$ degrees of
        freedom on $1792$ cores. We demonstrate the versatility of this
        framework with several examples, including sedimentation of particle
        clusters, and active matter systems composed of ensembles of particles
        driven to rotate.
        %%%%
    \end{abstract}

\end{frontmatter}

%%%%%%%%%%%%%%%%%%%%%%%%%%%%%%%%%%%%%%%%%%%%%%%%%%%%%%%%%%
%  main text 
%%%%%%%%%%%%%%%%%%%%%%%%%%%%%%%%%%%%%%%%%%%%%%%%%%%%%%%%%%

\section{Introduction}
\label{sec:intro}
% (1) What are particulate flows, why are they important, why should we develop methods to resolve collisions-----------------
Particulate suspensions are important to both technology and
fundamental science. The presence of particles suspended in fluid
leads to rich rheological behaviors of the mixture
\cite{Wagner_Brady_2009}, and novel applications.  For example,
shear-thickening colloidal fluids have been used to reinforce the
Kevlar-woven fabrics \cite{Lee_Wetzel_Wagner_2003} in bulletproof
vests.  Colloidal suspensions are important model systems in the study
of jamming \cite{Trappe_Prasad_Cipelletti_Segre_Weitz_2001} and phase
transitions \cite{Anderson_Lekkerkerker_2002}. Active suspensions
\cite{SS2013} are prototypes of active matter
\cite{MarchettiEtAl2013}, where the immersed particles may be
self-propelled, driven to rotate, or coupled together by biologically
active cross-linkers \cite{Shelley2016,ND2017}.

Here we discuss simulating the dynamics of particulate suspensions in
a Newtonian solvent. The accurate predictions of the dynamic
properties of such systems are key to understanding their behaviors
and to conceiving new applications. However, simulating such systems
is difficult. First, the intrinsic length scales in Stokes suspensions
can be large due to long-range many-body hydrodynamic interactions
(HIs hereafter).  Second, boundary conditions (no-slip, slip,
electrostatic, magnetic, etc.) on each particle must be accurately
satisfied. Otherwise, for example, the osmotic pressure measured from
dynamic simulations may show significant deviations from its true
value due to the numerical error in resolving particle-particle
collisions \cite{Foss_Brady_2000}. Third, such mixture systems must
often be tracked for long times due to slow relaxation processes or
significant Brownian noise in the system.

% (2) Approximate methods such as Stokesian dynamics, RPY, etc have been successful in scaling up the number of particles. However, full HD simulations of EHD/MHD/squirmers are all limited to isolated particles (cite -- Ishikawa, Biswal, etc). One of the key impediments is instabilities due to contact 
Many numerical methods have been developed to simulate suspensions in
a Newtonian solvent. The Rotne-Prager-Yamakawa tensor
\cite{Rotne_Prager_1969,Yamakawa_1970,Zuk_Wajnryb_Mizerski_Szymczak_2014,Wajnryb_Mizerski_Zuk_Szymczak_2013,Mizerski_Wajnryb_Zuk_Szymczak_2014,Beenakker_1986,Liang_Gimbutas_Greengard_Huang_Jiang_2013,guan2018}
is a popular approximation to account for HIs.  This method keeps the
mobility matrix $\bMcal$ symmetric-positive-definite (SPD), but is
rather crude for dense colloidal suspensions.  A more accurate method
is Stokesian Dynamics
\cite{Durlofsky_Brady_Bossis_1987,Brady_Bossis_1988,Phung_Brady_Bossis_1996,Sierou_Brady_2001,wang2016spectral},
which splits the HIs into a far-field and near-field part.  The
far-field part represents the HIs through multipole expansions
truncated at the stresslet level.  The near-field lubrication effects
are then added pairwise between particles with asymptotic lubrication
resistance functions.  Stokesian Dynamics has also been extended to
electrorheological suspensions
\cite{Bonnecaze_Brady_1992a,Bonnecaze_Brady_1992b} and non-spherical
particles \cite{Claeys_Brady_1993}.  However, it is difficult to
improve the accuracy of Stokesian Dynamics further due to the
algebraic complexity of including higher force moments on particles
beyond the stresslet level.

% (3) What is the current state-of-the-art and what are its shortcomings (force-penalty -- ts small, etc)---------
In all such computational methods, the efficient handling of
collisions between particles remains a long-standing problem.
Theoretically, for smooth rigid particles with no-slip boundary
conditions and moving under finite forces, lubrication effects
prevents their collision \cite{Kim_Karrila_2005}.  However, collisions
are inevitable in simulations because numerically it is impossible to
fully resolve the opposing lubrication effects with the finite
time-step sizes and finite accuracy of the HI solvers. Further,
particles with surface roughness or surface slip velocity may actually
collide. This feature is believed to underly the observed loss of flow
reversibility in the shearing of non-Brownian, ``spherical'' particle
suspensions \cite{PGBL2005}. In simulations, such collisions are
usually handled, or resolved, by using a prescribed short-range,
pairwise, repulsive potential between particles. For example, an
exponentially decaying pairwise repulsive force is used in Stokesian
Dynamics. Such pairwise potentials must be steep and therefore imposes
a restrictive upper bound on the time-step size to maintain
stability. Penalty methods \cite{janela2005} work similarly and suffer
from the same stability constraints. \footnote{In one interpretation,
    collision resolution algorithms are also models for the irreversible
    processes associated with the collisions of physical, and hence
    rough, particles.}
% Moreover, there is no guarantee that particles won't overlap no matter how stiff the penalty or the potential is, as numerically they cannot be infinite. 

% (4) Motivate why complementarity approach is attractive in this setting. Give a history of the method 
To build a stable and efficient numerical scheme, we reconsider the
collision resolution method. Pairwise potentials and penalty methods
compute collision forces on each particle at the start of each
time-step. Alternatively, collision forces can be solved for by
imposing non-overlapping constraints between bodies on particle
configurations over time. In this way, the stability issues induced
by the stiffness induced by potentials or penalties can be avoided.
\citet{Foss_Brady_2000} applied a `potential-free' algorithm in this
fashion, where overlaps are resolved by iteratively moving each
overlapping pair of particles apart until a non-overlapping
configuration is achieved.  This scheme, however, may converge slowly
in dense systems and does not allow many-body HI coupling between
particles.  Another method is the constrained minimization scheme
developed by \citet{Maury_2006}.  This method works as a
prediction-correction scheme, where a predicted velocity $\bU_p$ for
each particle is first computed, and then a corrected velocity $\bU$
satisfying the no-overlap constraints is obtained by minimizing the
$L_2$ norm of the difference $\bU-\bU_p$ for overdamped systems.  For
underdamped systems, the objective function being optimized is
modified to include the effects of mass and acceleration.  This
scheme, however, does not determine collision forces between each
colliding pair, making it difficult to compute the mechanical stress
induced by collisions.

Non-overlapping constraints can be coupled with collision forces in
inelastic collisions to construct complementarity constraints, since
each close pair of particles must be at one of the two possible
states: (1) they have collided, and so the collision force is positive
and their minimal separation is zero, or (2) they did not collide, and
so the collision force is zero and the minimal separation is positive.
The early development of these methods formulated collision resolution
problems in rigid body dynamics in various forms and probed their
mathematical properties, for example, the existence of solutions
\cite{cundall1979discrete,lotstedt1982,chunshengcai1987,montana1988,baraff1993}.
A linear complementarity formulation was soon developed to practically
simulate collections of frictionless \cite{anitescu1996} and
frictional \cite{stewart1996lcp,anitescu1997} rigid particles.  This
formulation uses a first-order Euler temporal discretization and
solves linear complementarity problems at each time-step.  This method
has been proved \cite{stewart1998,anitescu1999} to generate convergent
particle trajectories as the time-step $\Delta t \to 0$.  These early
developments have been summarized by \citet{stewart2000}, and extended
to handle stiff external forces \cite{anitescu2002} and bilateral
constraints such as mechanical joints \cite{anitescu2004}.  This early
formulation, however, considered non-convex linear complementarity
formulations for frictional particles and has been superseded by the
modern formulation of \citet{anitescu2006}, where a convex cone
quadratic program is solved at each time-step.  This convexity allows
efficent numerical solution in large scale simulations for frictional
granular flow problems
\cite{tasora2008,tasora2009,tasora2010,tasora2011,tasora2011a,negrut2012,tasora2013},
and open source software implementations of this method are now available \cite{heyn2013a,mazhar2013}.
Recently, this formulation has been
extended to deformable particles \cite{tasora2013a,pazouki2017}.  To
speed up these large scale simulations, various iterative solvers have
been discussed and compared in \cite{anitescu2010,heyn2013,mazhar2015,melanz2017,corona_RBDTT_2018}.
Recent work by two of the co-authors (Corona and Veerapaneni) and their co-workers demonstrated the
scalability of these methods on large distributed memory machines,
simulating the frictional multibody dynamics of hundreds of millions of granular particles \cite{De2019}.

For particulate Stokes flow problems, the recent work of \citet{lu_contact-aware_2017}
applied a similar explicit contact constraint enforcement approach for simulating deformable particles.
Furthermore, this approach was extended to three dimensional problems in \cite{lu_2018parallel}.
Both works demonstrated impressive gains in computational efficiency by
improving the stability and accuracy of the underlying numerical solvers.
However, we note that their implementation is tightly
bound to a particular boundary integral (BI) fluid solver.  Moreover,
the computed collision forces do not follow Newton's Third Law where
collision forces for a colliding pair must be equal and opposite.
Therefore, although the collision forces are explicitly computed in
this method, the collision stress may still be incorrect.

A proper extension of the complementarity method to Stokes particulate
suspensions is difficult because the particle motion is overdamped in
contrast to granular flows dominated by inertia. This requires
significant reformulation of the complementarity problem.  The solver
must also be improved because the complementarity problem now involves
a full dense matrix due to the many-body hydrodynamic coupling in
Stokes suspensions.  \citet{Yan_spherocylinder_2019} presented a
method to resolve normal collisions for arbitrarily-shaped rigid
particles in Stokes suspensions, as an extension of
\cite{anitescu1996}.  The complementarity problem for collision
resolution is reformulated and a different but more efficient solution
algorithm is described, utilizing the symmetric-positive-definiteness
of the mobility matrix.  This approach is generic because the motion
induced by collision force is represented by this mobility matrix, and
any hydrodynamic solver can be used.  This method imposes Newton's
Third Law and guarantees the symmetry and translational invariance of
the collision stress tensor.  They validated this method by computing
the equation of state for Brownian spherocylinders, and demonstrated
its application to various self-propelled rod systems.

Although the method is generic, results presented in
\cite{Yan_spherocylinder_2019} neglected the many-body HI coupling.
In this work, we extend this generic method to Stokes particulate
suspensions with full many-body HIs.  We use BI methods specialized for
spheres and achieving spectral accuracy by spherical harmonic
expansions for singular and near-singular integration
\cite{Corona_Veerapaneni_2018}.  The implementation is fully
parallelized with hybrid \texttt{OpenMP} and \texttt{MPI} to achieve
scalability.  We demonstrate the application of this platform to
various problems, such as sedimentation of particle ensembles, and the
collective dynamics of rotor systems.

The latter class of many-particle systems has been studied of late as
an active matter system, showing aspects such as activity-induced
phase separation \cite{yeoCollectiveDynamicsBinary2015},
crystallization \cite{OSS2019}, odd rheological and surface flow
dynamics \cite{SoniEtAl2019}, and forms of active turbulence
\cite{kokotActiveTurbulenceGas2017}. Motivated by the simulations of
\cite{yeoCollectiveDynamicsBinary2015}, \pr{fig:intro} shows a
large-scale simulation of $20,000$ particles, densely packed and
suspended on a plane in the fluid, with each particle driven by an
external out-of-plane torque. This simulation has roughly
$\sctf{6}{6}$ degrees of freedom, and was run on 576 CPU cores. The
simulation shows the development of an extensive and inhomogeneous
fine-scaled collision network, as well as the development of
large-scale collective rotation induced by long-ranged hydrodynamic
coupling. Of scientific interest is the interaction of multiple
such ensembles, and the detailed structure of particle flows within
them.
% We report accuracy and performance results for both the collision resolution algorithm and the BI mobility solver with examples including standard lubrication and sedimentation problems. 
% Strong and weak scaling benchmarks up to $1792$ CPU cores are demonstrated.
% We also demonstrate the application of this computational platform on Stokes rotors, which are spherical particles in Stokes flow driven by external torques applied on each particle.
% \pr{fig:intro} shows a large-scale simulation of $\sctf{2}{4}$ rotors with roughly $\sctf{6}{6}$ degrees of freedom on 576 CPU cores. 

\begin{figure}[ht!]
    \centering
    \includegraphics[width=\linewidth]{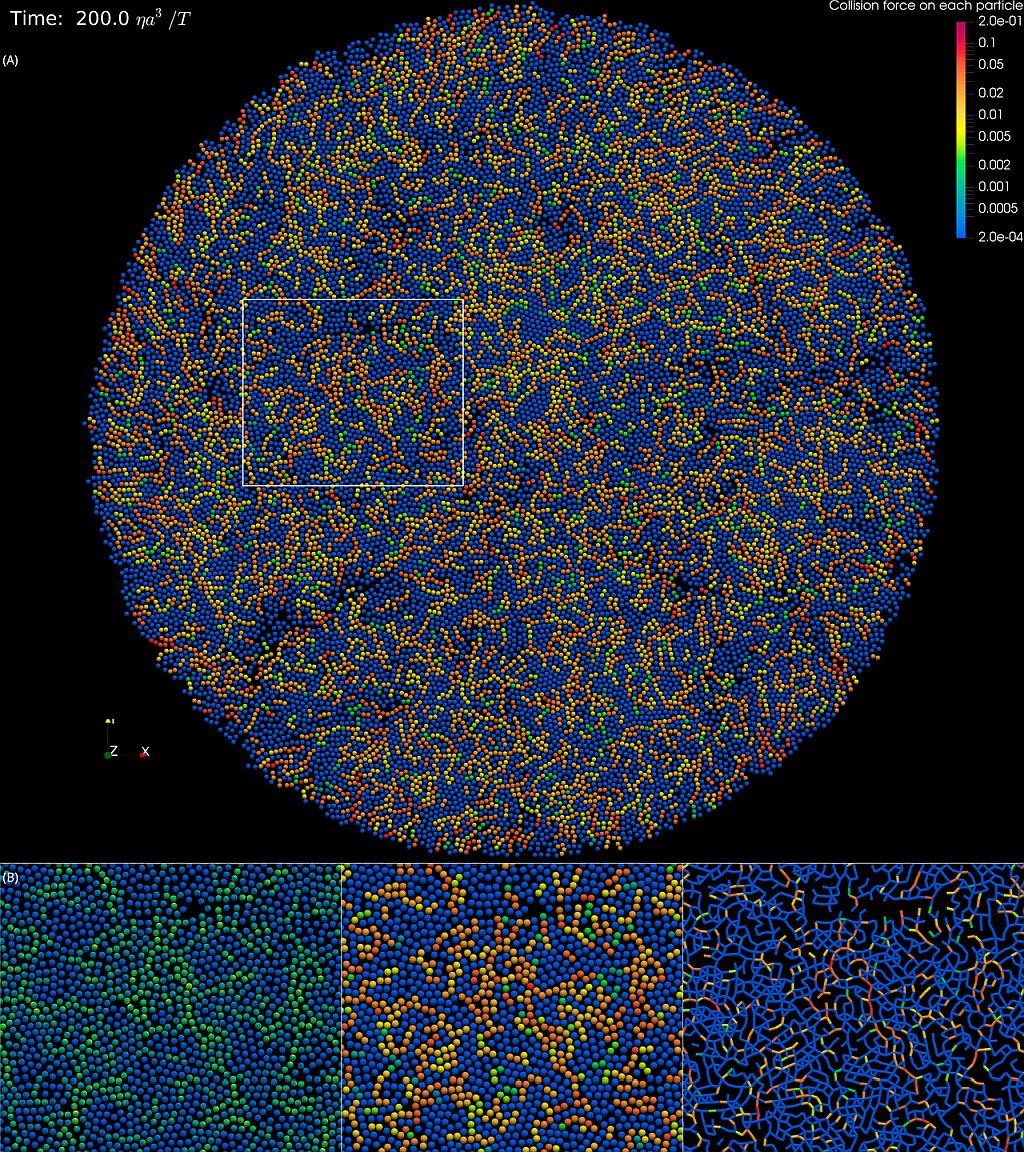}
    \caption{
        \label{fig:intro}
        A simulation snapshot of a monolayer of $20000$ spherical particles with radii $a$ at approximately $60\%$ area fraction.
        A constant external torque $\bT$ in the direction perpendicular to this monolayer is exerted on each particle to drive a counter-clockwise planar rotation of this mobolayer.
        (A) shows the entire simulation, with each particle colored by its total collision force magnitude.
        (B) shows the region inside the white box.
        The left panel of (B) shows the hydrodynamic force density (traction) on the sphere surfaces induced by collisions, scaled by $T/a^3$.
        The center and right panels show the net collision forces, scaled by $T/a$, on each particle and on each collision constraint, respectively.
        The three panels of (B) are colored by the same colormap as in (A).
    }
\end{figure}

\newpage
\section{Problem formulation}
\label{sec:problem}
% standard mobility problem
In this section we briefly summarize the well-known mobility problem,
i.e., the motion of arbitrarily-shaped particles driven by forces and
torques and immersed in a 3D Newtonian Stokes flow. Here, for
simplicity, the fluid occupies all space but the particle
volumes. 
Hence, consider a suspension of $N$ rigid particles, like
that shown in \pr{fig:intro} but not limited to spherical shapes. 
Let $\{V_j, \Gamma_j, \bc_j\}_{j=1}^N$ denote the volume, boundary, and
tracking points of the particles, respectively.
% {\bf MJS: This is  very confusing. 
% Geometric center sounds too much like center of   geometry which is the surface position average. 
% Plus, as you say, it is no center. So what do you want to call it? WY: replaced all `center' by `tracking point'.}  
Here, the `tracking point' refers to any point on one particle, not limited to its center-of-mass, because in this manuscript we consider only overdamped
dynamics of particles where inertia can be neglected.  
Each particle moves with a translational velocity $\bU_{j,e}$ and angular
velocity $\bOmega_{j,e}$, in response to an externally applied body
force $\bF_{j,e}$ and a body torque $\bT_{j,e}$.

Denoting the fluid viscosity by $\eta$, the fluid velocity by $\bu$,
and the pressure by $p$, the fluid stress is $\bsigma = -p \bI + \eta
    \left[\nabla \bu + \left(\nabla\bu\right)^T\right]$. Neglecting
inertial and body forces gives that $\nabla\cdot\bsigma = 0$ in the
fluid domain. Assuming fluid incompressibility, this gives the Stokes
equations:
\begin{subequations}
    \begin{align}
        -\nabla p + \eta \nabla^2 \bu & = 0 \quad\text{in}\quad \mbbR^3 \setminus \cup_j V_j,  \\
        \nabla \cdot \bu              & = 0  \quad\text{in}\quad \mbbR^3 \setminus \cup_j V_j, \\
        \bu                           & \to 0 \quad \text{as}\quad \bx\to\infty.
    \end{align}
    \label{eq:StokesEQ}
\end{subequations}

The traction $\bff$, i.e., the hydrodynamic force density on the particle surface applied by the fluid, is determined by the stress tensor $\bsigma$ as $\bff=\bsigma\cdot\bn$, where the surface normal $\bn$ points into the fluid domain.
The solution $(\bu_e,p_e)$ to \pr{eq:StokesEQ}(a-c) is subject to the boundary conditions:
\begin{subequations}
    \begin{align}
        \bu_e                                             & = \bU_{j,e} + \bOmega_{j,e} \times (\bx - \bc_j) \quad\text{on}\quad\Gamma_j,\forall\,  j, \\
        \int_{\Gamma_j} \bff_e \, dS                      & = \bF_{j,e},\quad\forall \, j,                                                             \\
        \int_{\Gamma_j} (\bx - \bc_j) \times \bff_e \, dS & = \bT_{j,e}, \quad \forall \, j .
    \end{align}
    \label{eq:extFTbc}
\end{subequations}%

There may also be collision a force $\bF_{j,c}$ and torque $\bT_{j,c}$
on each particle $j$ applied by other particles.  The consequent
velocities $(\bU_{j,c}, \bOmega_{j,c})$ induced by $\bF_{j,c}$ and
$\bT_{j,c}$ satisfy the same Stokes equation (\ref{eq:StokesEQ}) and
boundary conditions:
\begin{subequations}
    \begin{align}
        \bu_{c}                                             & = \bU_{j,c} + \bOmega_{j,c} \times (\bx - \bc_j) \quad\text{on}\quad\Gamma_j,\forall \, j, \\
        \int_{\Gamma_j} \bff_{c} \, dS                      & = \bF_{j,c},\quad\forall \, j,                                                             \\
        \int_{\Gamma_j} (\bx - \bc_j) \times \bff_{c} \, dS & = \bT_{j,c},\quad \forall \, j.
    \end{align} \label{eq:colFTbc}
\end{subequations}%

Due to the linearity of the Stokes equation (\ref{eq:StokesEQ}) and boundary conditions Eqs.~(\ref{eq:extFTbc}) and (\ref{eq:colFTbc}), the overall solution is simply the superposition of the two separate mobility problems induced by external and collisional forces and torques:
\begin{align}
    \bu=\bu_{e}+\bu_{c},\quad p=p_{e}+p_{c},\quad\bff=\bff_{e}+\bff_{c},\quad
    \bU_j=\bU_{j,e}+\bU_{j,c},\quad \bOmega_j=\bOmega_{j,e}+\bOmega_{j,c}.
    \label{eq:linearcomb}
\end{align}

Conventionally, the solution to a mobility problem can be written compactly as $\bUcal=\bMcal\bFcal$,
where \\ $\bUcal=\left(\ldots, U^x_j, U^y_j, U^z_j, \Omega^x_j, \Omega^y_j, \Omega^z_j, \ldots\right)$ and $\bFcal=\left(\ldots, F^x_j, F^y_j, F^z_j, T^x_j, T^y_j, T^z_j, \ldots\right)$ are both column vectors with 6 degrees of freedom per particle.
$\bMcal$ is the \emph{mobility matrix}, which is a dense square matrix containing all the information of the Stokes equation and boundary conditions.
The solutions given by \pr{eq:extFTbc} and \pr{eq:colFTbc} can be written compactly as $\bUcal_e = \bMcal\bFcal_e$ and $\bUcal_c = \bMcal\bFcal_c$, respectively.

$\bMcal$ is usually not formed explicitly because of the difficulty and high cost of computing its entries, except in cases where the many-body coupling of HIs is treated with very crude approximations \cite{Rotne_Prager_1969,Yamakawa_1970}.
Instead, a linear system $\bA\bx = \bb$ is usually constructed to solve the mobility problem.
The linear operator $\bA$ is usually computed from the geometry and boundary conditions, containing all the information of $\bMcal$.
The right hand side vector $\bb$ is usually computed from $\bFcal$, and the  velocities $\bUcal$ are computed according to the solution $\bx$.
The linear system $\bA\bx=\bb$ is ususally solved iteratively.

There are a variety of methods to construct the linear system $\bA\bx=\bb$.
Stokesian Dynamics \cite{Brady_Bossis_1988, Claeys_Brady_1993,Sierou_Brady_2001} forms the linear system using multipole expansions for both spherical and non-spherical particles.
BI methods forms the linear system using boundary integral operators and discretization of particle surfaces \cite{power1987second}.
When high levels of accuracy are necessary, BI methods provide the most accurate and efficient solvers to the mobility problem;
for example, see \cite{Corona_Greengard_Rachh_Veerapaneni_2017,Corona_Veerapaneni_2018,tornberg_numerical_2006,gustavsson_gravity_2009} and references therein.

% property of M
It is well-known that $\bMcal$ is symmetric-positive-definite
\cite{Durlofsky_Brady_Bossis_1987,Kim_Karrila_2005}.  Physically, this
is related to the dissipative nature of a Stokes suspensions.  The
energy injected by driving forces and torques is always dissipated
instantaneously by the fluid flow.  This is the key to our collision
resolution algorithm.

\section{Collision resolution algorithm}
\label{sec:formulation}
In this section we describe our collision resolution algorithm in
detail, making no assumptions on the particle shape, or on the
numerical method to solve the mobility problem $\bUcal=\bMcal\bFcal$.
The dynamics of particulate Stokes suspensions is overdamped, i.e.,
the particle inertia is negligible.  In this case, collisions are
inelastic with a zero coefficient of restitution.  Consequently, a
colliding pair of particles remain in contact until they are driven
apart by flow or collisions with other particles.  This is different
from granular flow, where inertial effects dominate and particles may
rebound after collisions, depending on their coefficients of
restitution.  We also ignore inter-particle friction for Stokes
suspensions since, physically, hydrodynamic lubrication effects
dominate for smooth particles close to contact.

\subsection{Complementarity formulation and equation of motion}
% part 1, continuous time DVI
In general, \pr{eq:linearcomb} can be generalized as:
\begin{align}
    \bUcal=\bUcal_{nc}+\bUcal_{c}.
\end{align}
Here the subscript $nc$ stands for all non-collisional motion, such as the motion $\bUcal_e$ driven by external forces and torques.
In other problems, $\bUcal_{nc}$ may originate from prescribed motions or other physical processes, such as Brownian fluctuations or electrophoresis.

% temporal equation of motion
The geometric configuration of a collection of rigid particles in 3D
space is fully specified by the tracking point location $\bc_j$ and
the orientation unit quaternion $\btheta_j=\{s,\bp\}\in\mbbR^4$ of
each particle $j$.  The temporal evolution of $\bc_j$ and $\btheta_j$
is given by:
\begin{align}
    \label{eq:sinparmotion}
    \dot{\bc_j}     & = \bU_j,                                                                                                             \\
    \dot{\btheta_j} & = \bPsi_j \bOmega_j, \quad \text{ where } \quad\bPsi_j = \frac{1}{2}\begin{bmatrix}
        -\bp_j^T \\
        s_j\bI - \bp_j
    \end{bmatrix} \in \mbbR^{4\times 3}.
\end{align}
Here we follow the rigid body kinematic equation using quaternions by \citet{delong2015}.
Similar to $\bFcal$ and $\bUcal$, we define the configuration $\bCcal$ as a column vector with $7N$ entries, containing $\bc_j$ and $\btheta_j$ for all N particles.
The overdamped equation of motion for these particles can be written compactly as:
\begin{align}
    \dot{\bCcal} & =  \bGcal \bUcal = \bGcal \bUcal_{nc} + \bGcal \bMcal \bFcal_{c}. \label{eq:pareom}
\end{align}
where $\bGcal\in \mbbR^{7N\times 6N}$ is a block diagnoal matrix, containing $3\times 3$ identity matrices and $4\times 3$ matrices $\bPsi_j$ corresponding to $\bc_j$ and $\btheta_j$ for each particle $j$, respectively.
Note that $\bGcal$ depends on those quaternion $\btheta$ components in $\bCcal$, but not on those $\bc$ components.
If the orientation of each particle is represented with Euler angles or rotation matrices, the kinematic equation can still be written in the form of \pr{eq:pareom}, but the dimension of $\bCcal$ and the definition of each $\Psi_j$ block must both be adjusted accordingly.
However, different representations of orientation does not affect the derivation of collision resolution algorithms \cite{anitescu1999}.

The equation of motion \pr{eq:pareom} should be augmented by geometric constraints to generate trajectories without overlaps between particles.
The constraints are simply that the minimal separation distance $\Phi_\ell(\bCcal)$ between each pair $\ell$ of close rigid objects remain non-negative for all configurations $\bCcal$.
In total there are potentially $n_c=N(N+1)/2$ such close particle-particle pairs.
For each pair $\ell$, we denote the collision force magnitude between this pair of particles as $\gamma_\ell$.
The pair $(\Phi_\ell, \gamma_\ell)$ must satisfy one of two conditions:
\begin{align}
    \text{No contact:}  \quad & \Phi_\ell>0,\quad\gamma_\ell=0. \\
    \text{In contact:}  \quad & \Phi_\ell=0,\quad\gamma_\ell>0.
\end{align}
Mathematically, this yields a \emph{complementarity problem}, written vertically over the set of pairs as:
\begin{align}
    \CP{\bPhi}{\bgamma}, 	\label{cp:cpproblem}
\end{align}
where $\bPhi=(\Phi_0,\Phi_1,...)\in\mathbb{R}^{n_c}$ and $\bgamma=(\gamma_0,\gamma_1,...)\in\mathbb{R}^{n_c}$ denote the collections of minimal distances and contact force magnitudes for each $\ell$.

For $N$ rigid particles, let $\bD_\ell \in \mathbb{R}^{6 N}$ be a sparse column vector mapping the magnitude $\gamma_\ell$ to the collision force (and torque) vector applied to each particle.
We define a sparse matrix $\bDcal\in\mathbb{R}^{6 N \times n_c}$ as a collection of all $\bD_\ell$:
\begin{align}\label{eq:Dmatdef}
    \bDcal & =\left[\bD_0\,\bD_1\,\ldots\,\bD_{n_c} \right] \in \mathbb{R}^{6 N \times n_c},\quad 	\bFcal_{c} =\bDcal\bgamma.
\end{align}
For each pair $\ell$, $\bD_\ell$ defined in this way has 12 non-zero entries for non-spherical shapes, corresponding to 3 translational and 3 rotational degrees of freedom for each particle.
For the special cases of two spheres, $\bD_\ell$ has 6 non-zero entries, as normal collision forces induce no torques on spheres.
The entries of each $\bD_\ell$ are explicitly given in \cite{anitescu1996}.
For rigid particles, there is an important relation between the transpose of $\bDcal$ and $\bPhi$ \cite{anitescu1996}:
\begin{align}\label{eq:Dtranspose}
    \bDcal^T =\left(\nabla_{\bCcal}\bPhi \right)\bGcal.
\end{align}
This relation holds for different choices of orientation
representations used in $\bCcal$ and $\bGcal$, including quaternions and
Euler angles \cite{anitescu1996}.

Combining \pr{eq:pareom}, \pr{cp:cpproblem}, and \pr{eq:Dmatdef}, we reach a differential variational inequality (DVI) for $N$ particles and $N(N+1)/2$ constraints.
\begin{subequations}\label{dvi:DVIdef}
    \begin{align}
        \dot{\bCcal} & = \bGcal \bUcal_{nc} +  \bGcal\bMcal \bDcal \bgamma, \\
                     & \CP{\bPhi(\bCcal)}{\bgamma}.
    \end{align}
\end{subequations}
Here $\bGcal$, $\bUcal_{nc}$, $\bMcal$, $\bDcal$, and $\bPhi$ depend only on the geometric configuration $\bCcal$.
This DVI is solvable and integrable over time once a relation between the configuration $\bCcal$ and the collision force $\bgamma$ is supplied, that is, a time-stepping scheme.
Higher order schemes such as the Runge-Kutta scheme can be used, but for simplicity of presentation we derive a first order Euler integration scheme here.

% part 2, time steps
\subsection{Time-Stepping and linearization}
With a given time-step $\Delta t$, the total number of constraints $n_c$ can be greatly reduced from $N(N+1)/2$ to $\Ocomp{N}$, since only those pairs that are close enough to be possibly in contact within this time-step need to be included in the constraints.
For example, for hexagonal close packing of spheres, every sphere is in contact with 12 neighbors.
This gives $n_c = 6N$.
We define $\Acal$ as the set of pairs of bodies `close-to-collision', that is, for which the minimal separation distance function $\Phi_{\ell}$ is smaller than a positive threshold $\delta$:
\begin{equation}
    \Acal(\bCcal,\delta) = \{ \ell \,\vert\, \Phi_\ell(\bCcal) \leq \delta \}.
\end{equation}
The choice of $\delta$ depends on the particle velocity $\bUcal$ and
the time-step $\Delta t$, and should be taken large enough so that no
possible collisions are missed in the collision resolution algorithm.
For each pair $\ell \in \Acal$, the non-overlap condition can be
simply stated as $\Phi_\ell(\bCcal) \geq 0$.  Increasing $\delta$
gives a larger set $\Acal$, and increases the dimension of the
\pr{cp:cpproblem}.  Empirically, we set $\delta$ to $30\%$ of the sum
of the radii for each pair of spherical particles.

The time-stepping scheme should evolve a non-overlapping configuration $\bCcal^k$ at $t^k$ to a non-overlapping $\bCcal^{k+1}$ at $t^{k+1}$.
Therefore, for the discretized version of the \pr{dvi:DVIdef}, the complementarity condition is between the geometry at the end of this time-step $\bPhi(\bCcal^{k+1})$ and the collision force at this time-step $\bgamma^k$:
\begin{align}
     & \bCcal^{k+1} = \bCcal^k + \bGcal^k \bUcal_{nc}^k \Delta t +  \bGcal^k \bMcal^k  \bDcal^k \bgamma^k \Delta t, \label{eq:time-step} \\
     & \CP{\bPhi\left(\bCcal^{k+1}\right)}{\bgamma^k}.
\end{align}
This is a nonlinear complementarity problem (NCP) because the minimum
separation $\bPhi$ is a nonlinear function of $\bCcal$.  For
first-order Euler time-stepping, this NCP can be linearized as an LCP
by taking a one-term Taylor expansion of $\bPhi$ over $\bCcal$:
\begin{align}\label{lcp:LCPderiv}
    \CP{\bPhi(\bCcal^{k}) + \left(\nabla_{\bCcal}\bPhi\right)^k\bGcal^k \bUcal_{nc}^k \Delta t + \left(\nabla_{\bCcal}\bPhi\right)^k\bGcal^k \bMcal^k \bDcal^k\bgamma^k \Delta t}{\bgamma^k}  .
\end{align}

At each time-step $t^k$ we solve the \pr{lcp:LCPderiv} to compute the
collision force magnitude $\bgamma$.  Then we use \pr{eq:time-step} to
update the geometry of all the particles. In summary, the procedures
of this LCP method include:
\begin{itemize}
    \item[1.] Compute $\bUcal_{nc}^k$ at time-step $k$.
    \item[2.] Construct the sparse matrix $\bDcal^k$ given the geometric configuration $\bCcal^k$ and apply threshold $\delta$ for possible contacts.
    \item[3.] Solve the LCP for $\bgamma^k$. $\bUcal_c^k$ and $\bFcal_c^k$ are computed simultaneously.
    \item[4.] Evolve the system to $\bCcal^{k+1}$ with $\bUcal_{nc}^k+\bUcal_c^k$.
\end{itemize}

This derivation is mostly the same as in the granular flow case \cite{anitescu1999}, except that inertia and friction are ignored.
However, this does not simplify the problem, because the mobility matrix $\bMcal$ is a full dense matrix, in contrast to the block-diagonal inertial and moment-of-inertia matrix in the granular flow case.
The high cost involved in computing $\bMcal\bDcal\bgamma$ requires careful attention to the solution strategy.
Once the solution to \pr{lcp:LCPderiv} is found, a subset $\bAcal_c$ of $\bAcal$ can be identified as the `active constraints', where the collision force between two particles is non zero:
\begin{align}
    \bAcal_c = \{ \ell \,\vert\, \gamma_\ell > 0 \}  \subseteq \bAcal.
\end{align}

\subsection{LCP solvers}
The success of the LCP collision resolution algorithm depends on the efficiency of the solver.
In this section we briefly discuss the state-of-the-art solvers used in this work.
The superscript $k$ denoting the time-steps are dropped in this subsection to simplify notation.

The LCP defined in \pr{lcp:LCPderiv} can be written as the following standard form, using \pr{eq:Dtranspose}:
\begin{subequations}\label{lcp:LCPdef}
    \begin{align}
         & \CP{\bM\bgamma+\bq}{\bgamma},                                                                       \\
         & \bM = \bDcal^{T} \bMcal\bDcal,\quad \bq = \frac{1}{\Delta t}\bPhi(\bCcal) + \bDcal^{T} \bUcal_{nc},
    \end{align}
\end{subequations}
where $\bM$ and $\bq$ have been scaled by $1/\Delta t$ for convenience.
The term $\bDcal^{T} \bUcal_{nc}$ computes the (linearized) change in the minimum separation function $\bPhi$ due to non-collisional motion.
Each evaluation of $\bM\bgamma$ corresponds to the solution of a mobility problem for the contact force $\bFcal_c = \bDcal\bgamma$.
For large enough numbers of particles, it may thus be preferable to rely on matrix-free operations instead of forming $\bM$ explicitly.

\subsubsection{Iterative solution methods}
The matrix $\bM$ defined in \pr{lcp:LCPdef} is in general symmetric-positive-semidefinite (SPSD), because the mobility matrix $\bMcal$ is SPD.
Therefore the LCP can be conveniently converted to a convex constrained quadratic programming (CQP) \cite{Niebe_Erleben_2015}:
\begin{equation}
    \bgamma = \arg\min_{\bgamma\geq 0}\left[ f(\bgamma) = \frac{1}{2} \bgamma^T \bM \bgamma + \bq^T \bgamma \right].\label{eq:CQPdef}
\end{equation}
For convenience, we denote the gradient of the objective function $f(\bgamma)$ as $\bg=\nabla f = \bM\bgamma+\bq$.
From a physics perspective, minimizing $f(\bgamma)$ can be understood as minimizing the total virtual work done by collision forces.
Popular methods to iteratively solve this CQP includes the second-order {minimum-map Newton} method, and the first-order projected gradient descent (PGD) method.

The minimum-map Newton method \cite{Niebe_Erleben_2015} proceeds by
reformulating the LCP as a root-finding problem for the $L_2$ norm of
the componentwise minimum-map function $\bH=\min(\bgamma,\bg)$, which means taking the smaller entry at every corresponding
component of the two vectors: $H_i=\min\{\gamma_i,g_i\}$. 
% {\bf MJS: $H_i=min\{\gamma_i,g_i\}$? WY: Yes}
\begin{align}\label{eq:mmapnorm}
    \varphi(\bgamma,\bg ) & = \NORMtwo{\min(\bgamma,\bg)}=0.
\end{align}
The solution $\bgamma$ to the CQP (\ref{eq:CQPdef}) is reached
when $\varphi=0$. 
% {\bf MJS: This is not a sentence. It begins
%   with an equation that doesn't really make sense. $\varphi$ sometimes
%   has two arguments, sometimes one. ?? WY: cleared this sentence and notation.} 
This Newton-type method can
achieve high accuracy, but the cost is usually much higher than first
order methods because a linear system must be iteratively solved at
every minimization step to compute the Newton step.

Various PGD methods have been proposed to solve this CQP efficiently because the gradient $\bg =\bM\bgamma+\bq$ is straightforward to compute, and the constraint $\bgamma\geq 0$ can be conveniently handled by setting the negative components of $\bgamma$ to $0$ because negative components violate the non-negative constraint of the LCP problem. This is a `projection' into the feasible region $\bgamma\geq0$, at each gradient descent step.
Such first-order methods do not find the root to \pr{eq:mmapnorm}.
Instead, \pr{eq:mmapnorm} is used to check the convergence with a prescribed residual tolerance $\epsilon_{tol}$.
When $\varphi(\bgamma, \bg) < \epsilon_{tol}$, the PGD iterations stop.

The key ingredient in all PGD methods is to choose a proper step-size
$\alpha_j$ for the $j$-th GD step: $\bgamma_{j+1} =
    \bgamma_{j}-\alpha_j\nabla f(\bgamma_{j})$.  The Accelerated PGD
(APGD) by \citet{mazhar2015} is shown to be a competitive choice in a
recent survey article \cite{melanz2017} in the context of granular
flow.  However, the step size $\alpha_j$ chosen by APGD converges only
when $\alpha_j < 1/L$, where $L$ is the Lipschitz constant for the
linear operator $\bM$ and is usually not known a priori.
Consequently, $\alpha_j$ is adaptively adjusted at each iteration to
fit a local estimated Lipschitz constant $L_j$ at the $j$-th step.
This adjustment process is prohibitively expensive in our context
since the gradient $\bg$ is evaluated a large number of times, each of
which requires solving a mobility problem.

A much more efficient method for \pr{eq:CQPdef} is the
Barzilai-Borwein PGD (BBPGD) method
\cite{Fletcher_2005,Dai_Fletcher_2005,Dai_Hager_Schittkowski_Zhang_2006}.
This method has been successfully applied to our collision resolution
method for cases where the many-body coupling of HIs are ignored
\cite{Yan_spherocylinder_2019}.  It constructs the step size
$\alpha_j$ from the previous two steps, and does not require any step
size adjustment as APGD.  A version of this method adapted to
\pr{eq:CQPdef} is summarized by \pr{alg:BBPGD}.  In this algorithm,
the two different choices of step size $\alpha_j^{BB1}$ and
$\alpha_j^{BB2}$ can be used either consistently throughout all steps,
or alternatively for odd and even $j$-th steps.  We find that there is
almost no difference in performance for different choices of
$\alpha_j^{BB1}$ or $\alpha_j^{BB2}$ (Step 15) in solving our
problems, and we use $\alpha_j^{BB1}$ for all results reported in this
work.  In BBPGD, two evaluations, $\bM\bgamma_0$ and $\bM\bg_0$ (Step
2 and 6), are necessary before the first iteration, and thereafter
only one evaluation $\bM\bgamma_j$ is needed per iteration.  In our
numerical tests BBPGD shows a similar convergence rate as APGD, but
each APGD iteration is significantly more expensive because of the
necessity to check the Lipschitz condition.

\begin{algorithm}[htbp]
    \begin{center}
        \begin{algorithmic}[1]
            \vspace{0.1in}
            \STATE Solve \pr{lcp:LCPdef} with initial guess $\bgamma_0$.
            \STATE $\bg_0=\bM\bgamma_0+\bq$.
            \IF{$\varphi(\bgamma_0,\bg_0)<\epsilon_{tol}$}
            \STATE Solution is $\bgamma_0$.
            \ENDIF
            \STATE Simple gradient-descent step size: $\alpha_0={\bg_0^T\bg_0}/{\bg_0^T\bM\bg_0}$.
            \FOR{ $j = 1 : j_{\max}$}
            \STATE Compute the descent step: $\bgamma_{j}=\bgamma_{j-1} - \alpha_{j-1}^{BB} \bg_{j-1}$.
            \STATE Projection: set the negative components of $\bgamma_j$ to zero. %$\bgamma_j=\Pi_{\bgamma\geq 0}\left[\bgamma_j\right]$.
            \STATE Compute the gradient: $\bg_j=\bM\bgamma_j+\bq$.
            \IF{$\varphi(\bgamma_j,\bg_j) < \epsilon_{tol}$}
            \STATE Stop iteration, solution is $\bgamma_j$.
            \ENDIF
            \STATE Update: $\bs_{j-1}=\bgamma_j-\bgamma_{j-1}$, $\by_{j-1}=\bg_j-\bg_{j-1}$.
            \STATE Update: $\alpha_j^{BB1}=\bs_{j-1}^T\bs_{j-1}/\bs_{j-1}^T\by_{j-1}$ or $\alpha_j^{BB2}=\bs_{j-1}^T\by_{j-1}/\by_{j-1}^T\by_{j-1}$. Set $\alpha_j^{BB}$ to either $\alpha_j^{BB1}$ or $\alpha_j^{BB2}$.
            \ENDFOR
        \end{algorithmic}
    \end{center}
    \caption{The Barzilai-Borwein Projected Gradient Descent method}
    \label{alg:BBPGD}
\end{algorithm}

\subsubsection{Choice of $\epsilon_{tol}$}
% {\bf MJS: I have no idea what the previous sentence is
%   trying to say. What dependencies are being discussed? This is, I
%   think, what provoked the referee's remark about trade-offs. WY: rephrased this paragraph.}  

The error bound for the general case SPSD matrix $\bM$ \cite{mangasarian1986} is highly complicated and there is no direct relation between the error bound to our residual function $\varphi$.
However, for most practical simulations the matrix $\bM$ is SPD.
This is because if we assume that there is a vector $\bgamma\neq\bm{0}$ such that $\bgamma^T\bM\bgamma=0$, then $\bDcal\bgamma=\bm{0}$ because $\bMcal$ is SPD in \pr{lcp:LCPdef}.
This means $\bDcal$ must be rank-deficient for its column vectors.
There exists at least one column vector $\bD_\ell$ linearly dependent on others.
Geometrically, this is possible but only for very special shapes and configurations of particles.
In most cases, the column vectors (constraints) in $\bDcal$ are independent of each other, and $\bM=\bDcal^T\bMcal\bDcal$ is SPD.

This allows us to use the simpler error bound for SPD matrices \cite{lin1987}. 
% {\bf as it is here, (?) WY: yes}
The absolute and relative error bounds for \pr{eq:CQPdef} are between an arbitrary vector $\bgamma$ and the exact
solution $\bgamma^*$ to the \pr{lcp:LCPdef}:

\begin{align}
    \NORMtwo{\bgamma-\bgamma^*}                             & \leq \frac{\NORMtwo{\bM}+1}{\lambda_{\min}} \varphi(\bgamma,\bg), \label{eq:abserrcqp}                                        \\
    \frac{\NORMtwo{\bgamma-\bgamma^*}}{\NORMtwo{\bgamma^*}} & \leq \mathrm{cond}(\bM) \left(\NORMtwo{\bM}+1\right) \frac{\varphi(\bgamma,\bg)}{\NORMtwo{\max(0,-\bq)}}.\label{eq:relerrcqp}
\end{align}

Here $\lambda_{\min}$ is the smallest eigenvalue of $\bM$,
$\NORMtwo{\bM}$ is the $L_2$-norm of $\bM$, $\mathrm{cond}(\bM)$ is the
condition number of $\bM$, and $\max(.,.)$ is the `componentwise
maximum map function', a counterpart to $\bH=\min(.,.)$.

%This allows us to use \pr{eq:abserrcqp} and \pr{eq:relerrcqp} to pick a reasonable $\epsilon_{tol}$.

However, $\lambda_{\min}$, $\NORMtwo{\bM}$, and $\mathrm{cond}(\bM)$
are all very difficult to estimate, because $\bM=\bDcal^T\bMcal\bDcal$
and only very crude estimates of $\bMcal$ are available
\cite{jiang2013a}: $\mathrm{cond}(\bMcal)\approx \Ocomp{1}$ of dilute
suspensions and $\mathrm{cond}(\bMcal)\approx \Ocomp{10^4}$ for
densely packed spherical particles.  A detailed error bound analysis
is beyond the scope of this work.  Instead, we pick $\epsilon_{tol}$
based on physical intuition and practical considerations.

We choose to set $\epsilon_{tol}$ as an absolute error bound,
following \pr{eq:abserrcqp}.  This is because the collision force for
different constraints can be orders of magnitude different, as
illustrated by the center panel of \pr{fig:intro} and in other results
reported in \pr{sec:results} and \pr{sec:application}.  
This is one of the physical features of rigid body suspensions, where the particle-particle interactions induced by perturbation may decay rapidly as the perturbation propagates and the particle-particle interaction in some regions of the system may be orders of magnitude larger or smaller than other regions. 
One example phenomena is Brinkman screening \cite{brinkman_calculation_1949} where the perturbation is `screened' hydrodynamically. 
If using the
relative bound, as in \pr{eq:relerrcqp}, the large collision forces
dominate the norm and the smaller entries in $\bgamma$ may be
completely inaccurate, leading to inaccurate trajectories or even
possibly overlapping configurations.  
We set $\epsilon_{tol}=10^{-5}$ for all results reported in this work unless otherwise specified.

In other works on granular flow \cite{heyn2013,mazhar2015}, there are
other residual functions used besides $\varphi(\bgamma,\bg)$. 
% {\bf MJS: Why is this a
%   residual.  What does it represent? What is $\Pi_c$? Is this its one
%   and only appearance? It seems to come out of nowhere... WY: removed the confusing residual function.} 
This is because in granular flow the matrix $\bM$ involves only a block
diagonal inertial and moment-of-inertia matrix, instead of a full
dense mobility matrix $\bMcal$. Various matrix splitting and
manipulation techniques can then be used to compute various residual
functions.  For our problem \pr{eq:CQPdef}, we stick to
$\varphi(\bgamma,\bg)$ because of its negligible extra cost of
computation at every gradient descent step, and because it is directly
related to the theoretical error bounds Eqs.~(\ref{eq:abserrcqp}) and
(\ref{eq:relerrcqp}).

\subsubsection{Extensions}
In the above discussion we constructed the basic form of our collision resolution algorithm.
There are two important extensions that can be easily incorporated.

The first straightforward extension is to include collisions between particles and external boundaries.
Here a `boundary' refers to any object either static or moving with a prescribed velocity.
In other words, a `boundary' does not appear in the mobility matrix $\bMcal$.
For each collision pair $\ell$ between a particle and a boundary, a vector $\bD_\ell$ can be added to $\bDcal$ in \pr{eq:Dmatdef}.
The only difference is that these $\bD_\ell$ column vectors have 6 non-zero entries corresponding to the degrees of freedom of the particle only.
Once $\bDcal$ is constructed including all these particle-boundary collisions, the solution to \pr{eq:CQPdef} remains exactly the same.

Another straightforward extension is to include particle motions other
than those driven by external forces.  For example, consider squirmers,
which are rigid particle models for ciliated organisms
\cite{Blake1971,IP2008}.
Squirmers propel themselves through a quiescent fluid,
without any external driving force, by inducing a nonzero surface slip
velocity $\bu_{j,slip}$ on the outer flow. For these particles, the
swimming motion $(\bU_{j,swim},\bOmega_{j,swim})$ can be computed by
solving the Stokes equation~(\ref{eq:StokesEQ}) subject to the
boundary conditions:
\begin{subequations}\label{eq:squirmerBC}
    \begin{align}
        \bu                                             & = \bu_{j,slip} + \bU_{j,swim} + \bOmega_{j,swim} \times (\bx - \bc_j) \quad\text{on}\quad\Gamma_j,\forall\,  j, \\
        \int_{\Gamma_j} \bff \, dS                      & = 0,\quad\forall \, j,                                                                                          \\
        \int_{\Gamma_j} (\bx - \bc_j) \times \bff \, dS & = 0, \quad \forall \, j.
    \end{align}
\end{subequations}
When external forces exist or squirmers may collide with each other,
this set of boundary conditions \pr{eq:squirmerBC} can be superposed
to the boundary conditions \pr{eq:extFTbc} and \pr{eq:colFTbc}.  The
swimming velocities $\bUcal_{swim}$ and externally driven velocities
$\bUcal_{e}$ can be computed straightforwardly and independently of each
other, without any knowledge about the collisional motion, at the
beginning of each time-step.  Then, we can apply the collision
resolution algorithm described in this section to compute the
collision velocities $\bUcal_c$, by simply setting the non-collisional
velocities as the sum of swimming velocities and externally driven
velocities:
\begin{align}
    \bUcal_{nc}=\bUcal_{swim} + \bUcal_e.
\end{align}

\section{Boundary integral formulation for the mobility problem}
\label{sec:bimobsolver}

The collision resolution method described above can be applied in conjunction with any mobility problem solvers for arbitrarily-shaped convex bodies.
Since one mobility problem needs to be solved at each evaluation $\bM\bgamma$ during the LCP solution, the computational cost and accuracy strongly depend on the mobility solver being used.

In this work, we apply a recently developed `indirect' BI formulation \cite{Corona_Greengard_Rachh_Veerapaneni_2017} to solve the mobility problem.
While the PDEs \pr{eq:StokesEQ} and \pr{eq:extFTbc} can be reformulated as an integral equation in a number of ways using potential theory \cite{pozrikidis1992boundary}, the advantage of this particular formulation is that second-kind integral equations are obtained without introducing any additional unknowns and constraints such as those in the work by \citet{power1987second}.
The discretized linear system can be solved rapidly with iterative solvers and, in most cases, $10 \sim 30$ iterations are sufficient irrespective of the problem size.

\subsection{Boundary integral formulation}
The reformulation of \pr{eq:StokesEQ} into a set of integral equations relies on the following standard operators:
\begin{align}
    \text{Single Layer Operator: } \SSg_\Gamma[\brho](\bx) & =\int_\Gamma \bG(\bx-\by) \, \brho(\by) \, d S_{\!\by},              \\
    \text{Double Layer Operator: } \SDb_\Gamma[\brho](\bx) & =\int_\Gamma \bT(\bx-\by) \,\bn(\by) \, \brho(\by) \, d S_{\!\by},   \\
    \text{Traction Operator: }  \STr_\Gamma[\brho](\bx)    & = \int_\Gamma   \bT(\bx-\by)\, \bn(\bx)\, \brho(\by) \, d S_{\!\by}.
\end{align}
Here $\bn$ is the outward surface normal vector to the rigid body
boundary $\Gamma$, and $\bG$ and $\bT$ are the fundamental solutions to
the Stokes equations. In particular, $\bG$ and $\bT$ are $2^{rd}$ and
$3^{nd}$ rank tensors known as the Stokeslet and traction kernels,
respectively, and are given by
\begin{align}
    G_{ij} (\br) & =\frac{1}{8\pi}\left(\frac{\delta_{ij}}{\ABS{\br}}+\frac{r_ir_j}{\ABS{\br}^3}\right), \\
    T_{ijk}(\br) & =-\frac{3}{4\pi}\frac{r_ir_jr_k}{\ABS{\br}^5}.                                        %=\left(G_{ij,k}+G_{ik,j}\right) - 2\frac{\delta_{jk}r_i}{\ABS{\br}^3}.
\end{align}
Note that the viscosity $\eta$ is set to unity here. For $N$ rigid
bodies suspended in free-space with respective boundaries
$\Gamma=\{\Gamma_j\}_{j=1}^N$, the indirect BI formulation of
\cite{Corona_Greengard_Rachh_Veerapaneni_2017} sets the fluid velocity
$\bu$ at an arbitrary point $\bx$ in the fluid domain as
\begin{equation}
    \bu(\bx) =   \SSg_{\Gamma}[\brho + \bzeta](\bx), \label{eq:velS}
\end{equation}
The unknown density functions $\brho$ and $\bzeta$ are determined as
follows. The role of $\brho$ is to match the given force and torque
(\pr{eq:extFTbc}d \& e) on each body.  It has been shown
\cite{rachh20152d, Corona_Greengard_Rachh_Veerapaneni_2017} that this
can be achieved by setting $\brho_j$ on each rigid body $j$ as:
\begin{align}
    \brho_j(\bx) = \frac{\bF_j}{|\Gamma_j|} + \btau_{j}^{-1}\bT_j\times \left(\bx-\bc_j\right) ,
\end{align}
where $\btau_j$ is the moment of inertia tensor
\cite{Corona_Greengard_Rachh_Veerapaneni_2017} and $|\Gamma_j|$ is the
surface area defined for particle $j$ with surface $\Gamma_j$.  For
the remaining unknown $\bzeta$, an integral equation can be obtained
by enforcing that the internal stress generated by $\brho+\bzeta$ on
each rigid body vanishes.  This is accomplished by:
\begin{align}
    \left(\frac{1}{2} \SId  + \STr_{\Gamma} + \SLo_\Gamma \right)[\bzeta](\bx) & = -\left(\frac{1}{2} \SId  + \STr_{\Gamma} \right)[\brho](\bx), \quad\forall \bx \in \Gamma, \label{eq:rigidbodyrho}
\end{align}
where for each $\Gamma_j$, $\SLo_{\Gamma_j}$ is a local operator
defined for each particle independently, designed to remove the 6 dimensional null space of the operator
$\frac{1}{2} \SId + \STr_{\Gamma}$: $\SLo_{\Gamma_j}[\bzeta](\bx)=\frac{1}{ \ABS{\Gamma_j} } \int_{\Gamma_{j}} \bzeta_{j}(\by) d S_{y}+{\btau}_{j}^{-1}\left(\int_{\Gamma_{j}}\left(\by-\bc_{j}\right) \times {\bzeta}_{j}(\by) d S_{y}\right) \times\left(\bx-\bc_{j}\right) $.  
We refer to \cite{Corona_Greengard_Rachh_Veerapaneni_2017} for more details.

In summary, the mobility problem is solved by first computing $\brho$
using given forces and torques, then solving \pr{eq:rigidbodyrho} for
$\bzeta$, and then computing the fluid velocity $\bu$ where needed
with \pr{eq:velS}.  The velocities $\bU_j,\bOmega_j$ for each rigid
body can be computed by averaging $\bu$ over the surface $\Gamma_j$.
In this approach, the operation $\bUcal = \bMcal \bFcal$ reqires an
iterative solution to \pr{eq:rigidbodyrho}, instead of a simple
explicit matrix-vector multiplication.

\subsection{Vectorial spherical harmonics discretization}
\label{ss:vec-sht-disc}
The BI formulation \pr{eq:velS} is applicable to arbitrarily-shaped rigid bodies, but the accuracy relies on the accurate evaluation of the operators $\STr$ and $\SSg$ for the given geometry.
For well-separated rigid bodies this is straightforward via various standard surface discretization and smooth quadrature rules.
However, for close pairs nearly in contact, special techniques must be employed 
because the Stokes kernels G and T become {\em nearly-singular}. 
While such techniques are well-developed for two-dimensional problems (see \cite{Wu2019} and references therein),
optimally handling arbitrary geometries in three dimensions is still an open problem 
and currently an active area of research \cite{siegel_local_2018,walaFastAlgorithmQuadrature2019,perez-arancibiaHarmonicDensityInterpolation2019}.
For spherical geometries, however, an efficient method based on vectorial spherical harmonic (VSH) basis functions was recently developed
in \cite{Corona_Veerapaneni_2018} by two of the co-authors. 
Here, we briefly summarize this VSH technique and apply it to develop a computational framework for spheres. 

% However for close pairs nearly in contact, special techniques must be employed because the Stokes kernels $\bG$ and $\bT$ are both singular.
% The Quadrature-by-Expansion method \cite{klockner2013,af_klinteberg_fast_2016-1,klinteberg_error_2017,siegel_local_2018,af_klinteberg_adaptive_2018,rahimian_ubiquitous_2018}, for example, is a general method for various shapes and kernels.
% For spheres, however, a specialized but much more efficient method based on the vectorial spherical harmonic (VSH) functions has been developed \cite{Corona_Veerapaneni_2018}.
% Here we briefly summarize this VSH technique and apply it to develop a computational framework for spheres.
% More details can be found in the work by \citet{Corona_Veerapaneni_2018}.

Any smooth vector field $\brho$, e.g., the hydrodynamic traction, defined on a spherical surface $\Gamma$ can be represented as an expansion over the VSH basis functions $\bV_n^m,\Wnm,\Xnm$:
\begin{align} \label{eq:rhoshexp}
    \brho = \sum_{n\geq0,-n\leq m\leq n} \left[\frac{1}{\ABS{\bV_n^m}^2} \hat{\rho}_{n,m}^V \bV_n^m + \frac{1}{\ABS{\Wnm}^2} \hat{\rho}_{n,m}^W \Wnm + \frac{1}{\ABS{\Xnm}^2} \hat{\rho}_{n,m}^X \Xnm \right],
\end{align}
where the basis functions $\Vnm,\Wnm,\Xnm$ are generated from the scalar spherical harmonic functions $\Ynm$:
\begin{align}
    \Gnm & =\nabla_\Gamma \Ynm, \\
    \Vnm & =\Gnm-(n+1)\Ynm \er, \\
    \Wnm & =\Gnm + n\Ynm\er,    \\
    \Xnm & =\er\times\Gnm,
\end{align}
where $\nabla_\Gamma= \dpone{ }{\theta}
    \et+\frac{1}{\sin\theta}\dpone{ }{\phi}\ep$ is the surface gradient
operator.  Note that when $n=0$, $\bV_0^0=-Y_0^0\er$ does not vanish,
being just a vector field pointing inward on the unit sphere surface.

The coefficients $\hat{\rho}_{n,m}^V$, $\hat{\rho}_{n,m}^W$,
$\hat{\rho}_{n,m}^X$ are the inner product of $\brho$ and the basis
functions on the spherical surface:
\begin{align}
    \hat{\rho}_{n,m}^V & = \AVE{\brho,\Vnm},\quad \hat{\rho}_{n,m}^W = \AVE{\brho,\Wnm},\quad \hat{\rho}_{n,m}^X = \AVE{\brho,\Xnm},
\end{align}
where
\begin{align}
    \AVE{\bu,\bv} & = \int_S \bu\CONJ{\bv} dS.
\end{align}
For real-valued $\brho$, the $n, m$ and $n, -m$ coefficients are complex conjugates to each other.
In this work we follow the notational convention for spherical harmonics, and write $\Ynm$ as
\begin{align}
    \Ynm (\theta,\phi) & = \sqrt{\frac{2n+1}{4\pi}} \sqrt{\frac{(n-m)!}{(n+m)!}}\Pnm (\cos\theta) e^{im\phi},
\end{align}
where
\begin{align}
    \Pnm (x) & = \frac{1}{2^n n!}(-1)^m (1-x^2)^{m/2} \frac{\partial^{n+m}}{\partial x^{n+m}} \left(x^2-1\right)^n,
\end{align}
are Legendre polynomials.
The $P_n^{\pm m}$ satisfy:
\begin{align}
    P_n^{-m} (x) = (-1)^m\frac{(n-m)!}{(n+m)!}\Pnm (x).
\end{align}

The expansion \pr{eq:rhoshexp} is spectrally convergent for a smooth density $\brho$.
Similarly, for a target point not on the spherical surface $\Gamma$, the value of the BI operators evaluated at this point $\SSg_\Gamma[\brho](\bx)$, $\SDb_\Gamma[\brho](\bx)$, and $\STr_\Gamma[\brho](\bx)$ can all be expanded as a series summation over the VSH basis.
For example, for a point $\bx$ outside $\Gamma$, we first write $\bx=(r,\theta,\phi)$ in the spherical coordinate system of $\Gamma$.
Then, the velocity $\bu(r,\theta,\phi) = \SSg_\Gamma[\brho](\bx)$ and the corresponding fluid pressure $p(r,\theta,\phi)$ governed by Stokes equation can be expressed by:
\begin{align}\label{eq:ssgshexp}
    \bu(r,\theta,\phi) & = \sum_{n,m} f_{n,m}^V(r)\Vnm + f_{n,m}^W(r)  \Wnm + f_{n,m}^X(r) \Xnm, \\
    p(r,\theta,\phi)   & =\sum_{n,m} g_{n,m}(r) \Ynm.
\end{align}
The mapping from the VSH coefficients for $\brho$ in \pr{eq:rhoshexp} to $f_{n,m}^V(r),f_{n,m}^W(r),f_{n,m}^X(r),g_{n,m}(r)$ is linear due to the linearity of Stokes equation, and is diagonal:
\begin{subequations}\label{eq:ssgext}
    \begin{align}
        f_{n,m}^V(r) & = \frac{n}{(2n+1)(2n+3)}r^{-(n+2)}  \frac{1}{\ABS{\bV_n^m}^2} \hat{\rho}_{n,m}^V  + \frac{n+1}{4n+2} \left({r^{-(n+2)}}-r^{-n}\right) \frac{1}{\ABS{\Wnm}^2} \hat{\rho}_{n,m}^W , \\
        f_{n,m}^W(r) & = \frac{n+1}{(2n+1)(2n-1)}r^{-n} \frac{1}{\ABS{\Wnm}^2} \hat{\rho}_{n,m}^W  ,                                                                                                     \\
        f_{n,m}^X(r) & = \frac{1}{2n+1}r^{-(n+1)} \frac{1}{\ABS{\Xnm}^2} \hat{\rho}_{n,m}^X,                                                                                                             \\
        g_{n,m}(r)   & = n r^{-(n+1)} \frac{1}{\ABS{\Wnm}^2} \hat{\rho}_{n,m}^W .
    \end{align}
\end{subequations}
Similar diagonalized relations have been derived by \citet{Corona_Veerapaneni_2018} for both $\SSg$ and $\SDb$ operators, and for $\bx$ both inside and outside $\Gamma$.

For the traction operator $\STr$, this mapping is no longer diagonal.
In other words, the mapping from the VSH coefficients for $\brho$ to the coefficients of $\STr_\Gamma[\brho](\bx)$ is a dense matrix.
In this work, we derive a general analytical relation, for the target point $\bx$ both inside and outside $\Gamma$, of this full dense mapping using \pr{eq:ssgshexp}.
The traction $\bt$ at each target point $\bx$ is defined as:
\begin{align}
    \bt & =\bm{\sigma}\cdot\bn=\left(-p\bI+\nabla \bu + \nabla \bu^T\right)\cdot\bn,
\end{align}
where $p$ has been given in Eq.~(\ref{eq:ssgshexp}) and (\ref{eq:ssgext}).
The velocity gradient tensor $\nabla\bu$ has 9 components in the $(r,\theta,\phi)$ spherical coordinate system:
\begin{align}\label{eq:tracgnm}
    \nabla \bu & =\sum_{n,m} \begin{bmatrix}
        g_{nm}^{rr}       &  & g_{nm}^{r\theta}      &  & g_{nm}^{r\phi}      \\
        g_{nm}^{\theta r} &  & g_{nm}^{\theta\theta} &  & g_{nm}^{\theta\phi} \\
        g_{nm}^{\phi r}   &  & g_{nm}^{\phi\theta}   &  & g_{nm}^{\phi\phi}   \\
    \end{bmatrix}.
\end{align}
Each $g_{nm}$ can be analytically computed with the functions $f_{nm}^V$, $f_{nm}^W$, $f_{nm}^X$ in \pr{eq:ssgext}.
The detailed expressions are given in \pr{appsec:traction}.
The velocity gradient $\nabla \bu$ is then transformed from the spherical coordinate system $r,\theta,\phi$ to the Cartesian coordinate system by standard tensor rotation rules.

For a smooth density $\brho$ defined on a spherical surface $\Gamma$, $\SSg_\Gamma[\brho](\bx)$, $\SDb_\Gamma[\brho](\bx)$, and $\STr_\Gamma[\brho](\bx)$ are again spectrally convergent.
This important feature allows us to represent lubrication effects efficiently with only a few spherical harmonic modes.
In simulations of many spheres, the operators $\SSg_\Gamma[\brho](\bx)$, $\SDb_\Gamma[\brho](\bx)$, and $\STr_\Gamma[\brho](\bx)$ are first directly evaluated on discretized spherical harmonics grid points \cite{Corona_Veerapaneni_2018} defined on the surface of each sphere by Kernel Independent Fast Multipole Method (KIFMM).
However, the results evaluated by KIFMM are inaccurate when spheres are close to each other, because the Stokes kernels $G_{ij}$ and $T_{ijk}$ are singular
To overcome this inaccuracy, for each close pair of spheres $i,j$ we first transform the density $\brho$ from the surface grid points to the VSH coefficients $\hat{\rho}_{n,m}^V$, $\hat{\rho}_{n,m}^W$, and $\hat{\rho}_{n,m}^V$.
Then, the VSH representation is used to evaluate the accurate values of the boundary integral operators $\SSg_\Gamma[\brho](\bx)$, $\SDb_\Gamma[\brho](\bx)$, and $\STr_\Gamma[\brho](\bx)$.
The close pairs are detected by checking the center-to-center distance $\ABS{\bc_i-\bc_j}$.
If this distance is smaller than a threshold value $\beta (R_i+R_j)$, the VSH representation is used.
In this work, we usually choose $\beta\in(1.5,2)$.
We observed no benefits in accuracy for larger $\beta$, and the cost of close-pair VSH corrections quickly increases with increasing $\beta$.

\section{Implementation}
\label{sec:implementation}
Proper implementation is necessary to achieve a scalable computational framework with modern \texttt{MPI+OpenMP} parallelism that maximizes the efficiency on high-core-count CPUs.
In this section we describe the four major components in our implementation.

\subsection{KIFMM}
The mobility problem is solved via GMRES iteration of the BI \pr{eq:rigidbodyrho}, where the operator $\STr$ must be evaluated once every GMRES iteration.
The operator $\STr$ is an all-to-all operator, where the traction kernel $\bT$ is evaluated between every pair of the spherical harmonic grid points on all spheres.
In total, there are $(p+1)(2p+1)N$ points for $N$ spheres with order $p$ spherical harmonics.
The operators $\SSg$ and $\SDb$ are also evaluated in this all-to-all style.
These are standard operations and can be computed by KIFMM with $\Ocomp{N}$ cost.
In this work, we use the fully parallelized KIFMM package PVFMM \cite{malhotra_pvfmm_2015}, and code the kernel functions as optimized \texttt{AVX2} intrinsic instructions to fully utilize the 256-bit \texttt{SIMD} capability of modern CPUs.
The developed code is open-sourced as \texttt{STKFMM} on GitHub.\footnote{\url{https://github.com/wenyan4work/STKFMM}}
We verified the accuracy of the operator evaluations to machine precision, and benchmarked the scalability to thousands of cores on a CPU cluster interconnected by Intel Omni-Path fabrics.

\subsection{Near neighbor detection}
Once the BI operators are evaluated with KIFMM, near-field corrections
must be performed for close-to-contact pairs of spheres with the VSH
representation.  This step requires efficient detection of all
close-to-contact pairs.  This is a standard neighbor detection
operation and can be efficiently completed with algorithms such as
cell list or k-D tree.  However, although there are a few
high-performance libraries publicly available such as FDPS
\cite{iwasawa_implementation_2016}, DataTransferKit
\cite{slattery_mesh-free_2016}, and LibGeoDecomp
\cite{schafer_libgeodecomp_2008}, there are some important features
still missing.  For example, the necessary VSH data, including grid
point coordinates, values, and expansion coefficients, needs to be
efficiently migrated between the \texttt{MPI} ranks, and customizable
serialization and de-serialization are necessary to allow VSH data
with different order $p$ to be sent and received as $\texttt{MPI}$
messages.

Therefore we built a custom near neighbor detection module based on a
Morton-coded octree.  This near neighbor module is fully parallelized
with both \texttt{OpenMP} and \texttt{MPI}.  Once the near neighbor
pairs are detected, the necessary data is transferred in
\texttt{msgpack} binary format with the \texttt{msgpack-c}
library.\footnote{\url{https://github.com/msgpack/msgpack-c}} Each
pair is dispatched to one \texttt{OpenMP} thread to compute the near
corrections, implemented in a thread-safe way.  This near neighbor
detection module is also used to identify the possibly colliding pairs
to construct the set for all constraints $\bAcal$.  Then the sparse
matrix $\bDcal$ is constructed for the \pr{lcp:LCPdef}, distributed on
all \texttt{MPI} ranks.

\subsection{Load-balancing}
Proper load balancing is necessary to ensure scalability over
\texttt{MPI} parallelism, and the balancing of both the LCP and the
mobility problem must be considered.  The balancing of the mobility
problem \pr{eq:rigidbodyrho} is handled through the KIFMM routine and
the near neighbor detection routine.  An adaptive octree with 2:1
balancing is built inside the PVFMM library and decomposed to each MPI
rank to allow efficient KIFMM evaluation.  Similar decomposition is
also used in the near neighbor detection module so that each
\texttt{MPI} rank handles roughly the same number of near neighbors
pairs to perform VSH corrections.  The decomposition for the LCP is
slightly different.  In \pr{lcp:LCPdef}, each contact pair appears
only once in the vector $\bgamma$ and the geometric matrix $\bDcal$.
We use a simple but effective strategy to partition $\bgamma$ and
$\bDcal$.  Each particle is labeled with an index $i$, which is
globally unique across all \texttt{MPI} ranks and randomly initialized
in the beginning of simulations, so that $i$ is uncorrelated with its
spatial location $\bc_i$.  When the pair of particles $i,j$ is
detected to be possibly colliding, the minimum separation $\Phi_{ij}$,
the sparse colliding geometry column vector $\bD_{ij}$, and the
unknown collision force $\gamma_{ij}$, are determined to be owned by
the $\texttt{MPI}$ rank which owns the smaller index between $i,j$.
The transposed matrix $\bDcal^T$ is then constructed in
compressed-row-storage (CRS) format where the rows are partitioned to
each \texttt{MPI} rank.  $\bPhi$ and $\bgamma$ are partitioned in the
same way as $\bDcal^T$.  Compared to the matrix-free implementation
where the matrix $\bDcal^T$ and $\bDcal$ are not explicitly
constructed \cite{tasora2011}, our implementation takes more storage
space but allows the utilization of the highly optimized sparse linear
algebra functions implemented in the libraries \texttt{Tpetra} and
\texttt{Kokkos} in
\texttt{Trilinos}\footnote{\url{https://trilinos.org}}.

\subsection{Vector spherical harmonics}
% fill by Dhairya
% transform, evaluation, BLAS, thread safety
For spherical harmonics of order $p$, we compute the VSH expansion
coefficients from function values on the $(p+1)(2p+1)$ spherical
harmonic grid points using a spherical harmonic transform (SHT).  We
do this by computing $2p+1$ discrete Fourier transforms (DFT) in the
$\phi$ direction and a Legendre transform in the $\theta$ direction
for each sphere.  The DFT is computed using the FFTW3 functions
\cite{FFTW3_2005} and requires $\Ocomp{p^2 \log p}$ work per sphere.
We compute the Legendre transform using matrix-vector products, and
this requires $\Ocomp{p^3}$ work per sphere.  We do not use a fast
Legendre transform (FLT) since it is advantageous only for very large
$p$.  For multiple spheres, we parallelize using OpenMP by
partitioning the spheres across threads.  When the number of spheres
is greater than the number of threads, we use blocking to compute
multiple transforms together.  This allows us to use matrix-matrix
products for the Legendre transform, and these are computed
efficiently using an optimized
BLAS\footnote{\url{http://www.netlib.org/blas}} implementation.  We
use Intel Math Kernel Library for both the FFTW3 and BLAS functions.
We can similarly compute function values on the spherical harmonic
grid from the spherical harmonic coefficients using the inverse SHT.
For vector valued functions, we can compute the coefficients for the
representation in the VSH basis by computing three scalar SHTs for
each sphere as discussed in \cite{Corona_Veerapaneni_2018}.
Similarly, we can compute the grid values from the VSH coefficients
using three inverse SHTs.

Once the VSH coefficients for the density function $\brho$ have been computed for each sphere, we can then compute the coefficients for the velocity $\bu$, the pressure $p$ and the traction $\bt$ for each sphere-target pair using the diagonal operators discussed in Section \pr{ss:vec-sht-disc}.
Finally, we evaluate the VSH expansions to get the potentials.
This requires evaluating the VSH basis functions and has $\Ocomp{p^2}$ computational cost for each sphere-target pair.
Since the positions of the spheres only change between time-steps, we could precompute these basis functions at each time step and reuse them during the linear solve.
However, this would require $\Ocomp{p^2}$ memory for each sphere and near target pair; therefore, we do not do this precomputation in our current implementation.

%Numerical Experiments 
\section{Results}
\label{sec:results}
In this section we report the numerical accuracy and performance results for suspensions of spherical particles in unbounded Stokes flow.

\subsection{Static lubrication benchmark}
In this section we probe the accuracy of the mobility solver discussed in \pr{sec:bimobsolver}, for a few static configurations.
The collision resolution algorithm is not used.

\subsubsection{Two spheres}

\begin{figure}[htbp]
    \centering
    \includegraphics[width=0.8\linewidth]{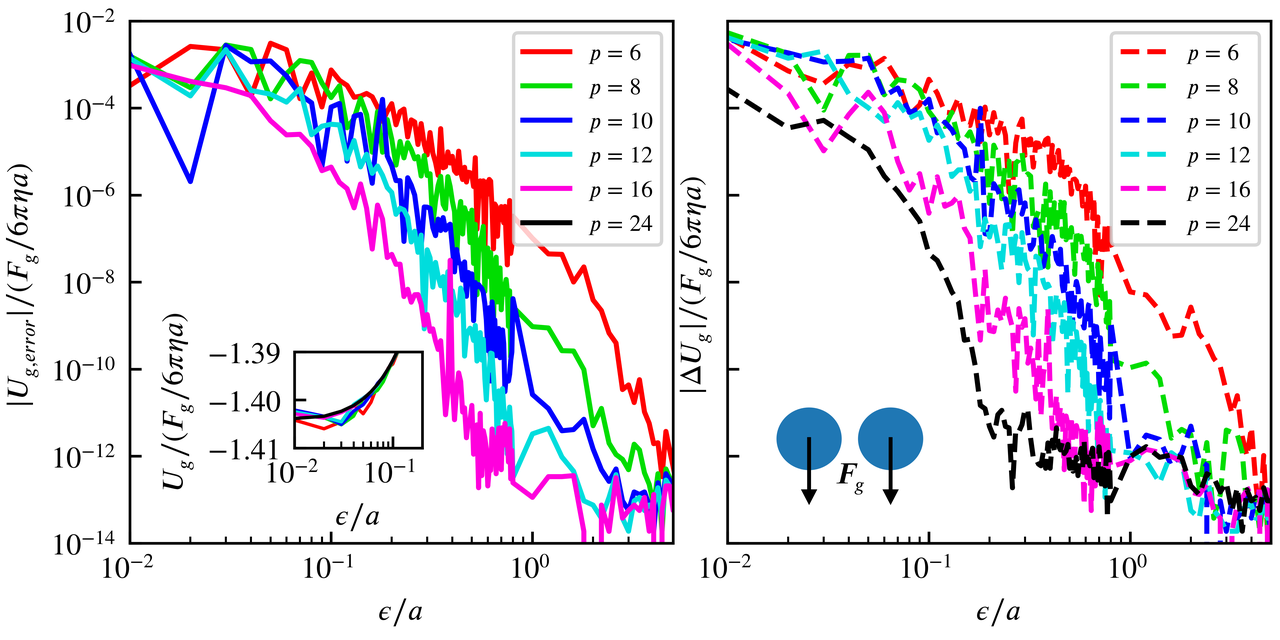}
    \caption{\label{fig:lubsedi} The convergence error and asymmetry
        error for the sedimentation velocity $\bU_g$. The inset plot
        shows the sedimentation velocity in comparison to the single
        sphere Stokes solution $\bF_g/(6\pi\eta a)$.}
\end{figure}

\begin{figure}[htbp]
    \centering
    \includegraphics[width=0.8\linewidth]{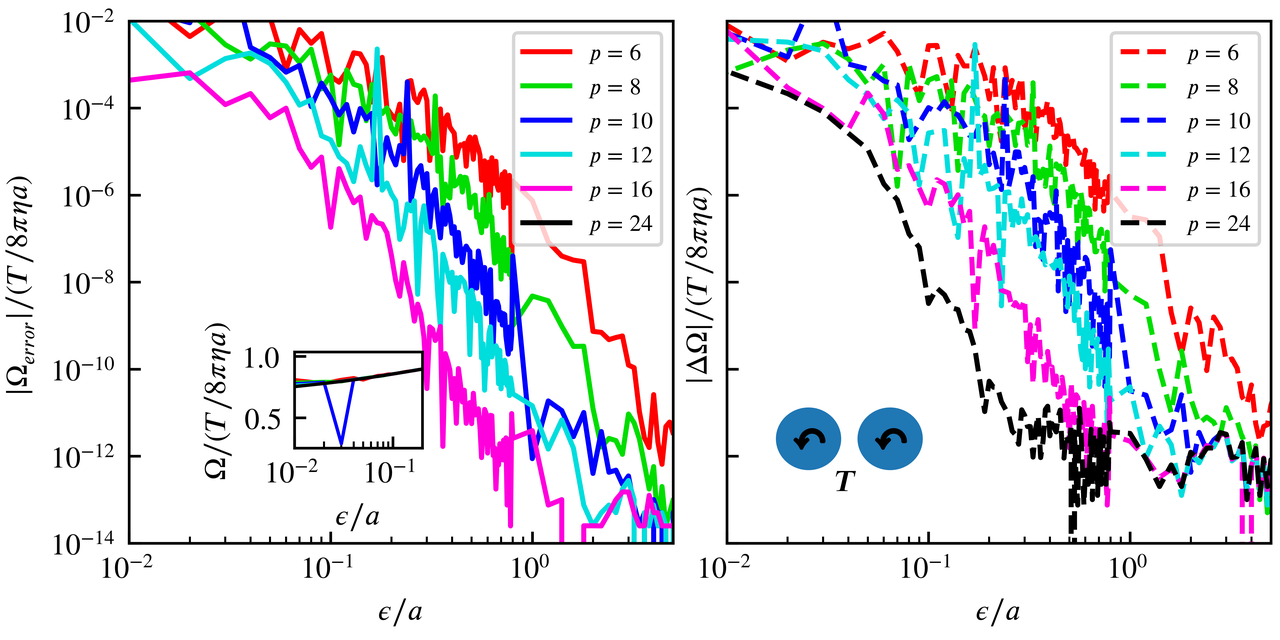}
    \caption{\label{fig:lubroll} The convergence error and asymmetry
        error for the angular velocity $\bOmega$. The inset
        plot shows the angular velocity in comparison to the single
        sphere Stokes solution $\bOmega = \bT/(8\pi\eta a^3)$. }
\end{figure}

We begin with a convergence test for a static configuration where
lubrication forces are important.  A pair of nearby spheres with
radius $a$ are driven by the same force $\bF_g$, i.e. sedimentation,
or torque $\bT$ as illustrated by the schematics in
Figs. \ref{fig:lubsedi} and \ref{fig:lubroll}, respectively.  To
evaluate convergence, $\bU$ and $\bOmega$ are solved for both spheres,
for each case, at various gap separations $\epsilon$.  The errors are
shown in \pr{fig:lubsedi} for the sedimentation case and in
\pr{fig:lubroll} for the rotation case.  In each figure, the left
panel shows the convergence error (absolute value) relative to the
results generated by $p=24$:
$\ABS{U_{g,error}(p)}=\ABS{U_g(p)-U_g(24)}$ and
$\ABS{\Omega_{error}}=\ABS{\Omega(p)-\Omega(24)}$, of the particle on
the left.

It is also well-known that spherical harmonic grids are significantly denser around the poles in comparison to areas close to the equator. 
This non-uniformity across the sphere surface may affect the capability to resolve the highly non-uniform distribution of hydrodynamic force $\bff$ induced by lubrication effects. 
Because of this, the spherical harmonics grid is randomly oriented for each sphere at different
$\epsilon$.  
This randomness induces some asymmetry error in the
computed $\bU$ and $\bOmega$, defined as the difference of computed
velocity for the two particles $\Delta U_g = U_{g,1}-U_{g,2}$ and
$\Delta\Omega=\Omega_{1}-\Omega_{2}$, as shown in the right panels of
Figs. \ref{fig:lubsedi} and \ref{fig:lubroll}.

The results show that the asymmetry error is always on the same order
as the convergence error.  
Therefore, the user does not need to pick a particular pole orientation to `better resolve' the lubrication effect. 
Further, spectral convergence holds until
the separation $\epsilon/a$ is comparable with $1/p$.  This is because
for such small gaps the hydrodynamic traction has a large peak in the
near-contact region due to lubrication effects.  Also, the error for
the rotation case is larger than for the sedimentation case, because
when the two spheres are rotating in the same direction, the fluid in
the near-contact region has a very large shear rate due to the
relative motion of the two sphere surfaces.

\subsubsection{Three spheres}
Accurate benchmark data for lubrication effects is surprisingly hard
to find because the dominating method in this setting was asymptotic
expansions \cite{Durlofsky_Brady_Bossis_1987,Kim_Karrila_2005} with questionable convergence beyond 3 digits of accuracy \cite{wilson_stokes_2013}.
To our knowledge the most accurate lubrication benchmark is given by
\citet{wilson_stokes_2013} to 10 digits of accuracy for a few
particular geometries of several spheres.  We use these results to
evaluate the accuracy of our mobility solver based on KIFMM and VSH
for three equidistant spheres of equal radii each driven by a constant
force $\bF$ perpendicular to their common plane.  The results for the
relative error of the translational velocity $\bU$ and angular
velocity $\bOmega$ are shown in \pr{fig:lubtriangle} as a function of
the gap separation distance $\epsilon$ between each pair of spheres.

\begin{figure}[htbp]
    \centering
    \includegraphics[width=0.8\linewidth]{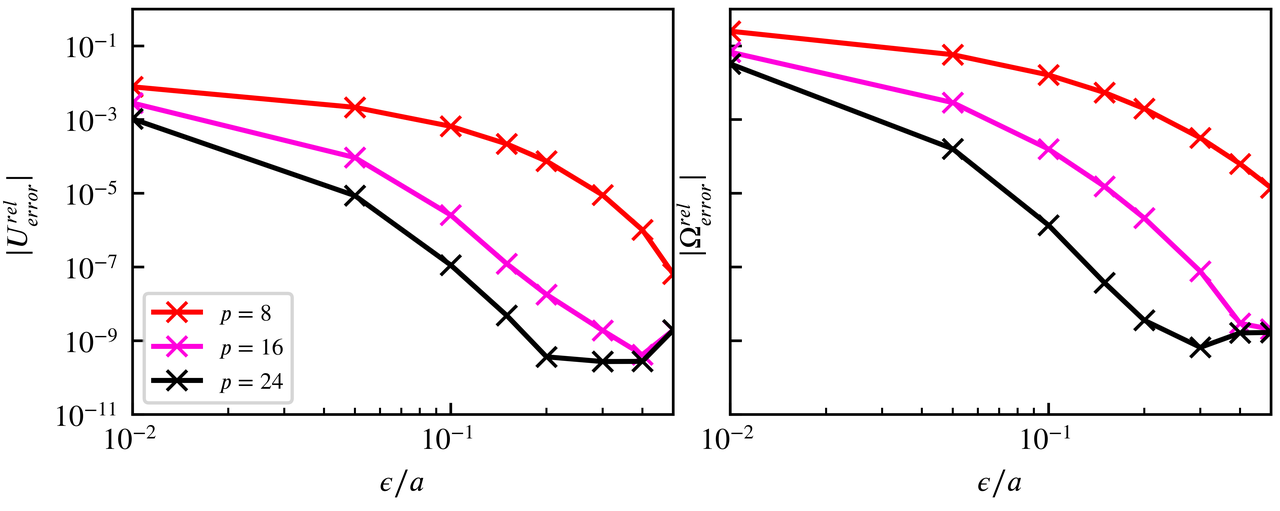}
    \caption{\label{fig:lubtriangle} The relative error of
        translational and angular velocity in comparison to high
        accuracy reference data \cite{wilson_stokes_2013}. The error is
        limited to $10^{-10}$ as this reference data has 10 digits of
        accuracy.}
\end{figure}

The results in \pr{fig:lubtriangle} show accuracy similar to the
two-particle benchmarks shown previously in \pr{fig:lubsedi} and
\pr{fig:lubroll}.  It is clear that the spectral convergence with
increasing $p$ is kept until $\epsilon/a$ is comparable to $1/p$,
which is roughly at $\epsilon/a\approx0.05$ for $p=24$.  When
$\epsilon$ further decreases, the error decreases slowly with
increasing $p$.

% compare with Stokesian Dynamics and justify the collision radius
% be very careful
Lubrication effects are added explicitly in the popular Stokesian Dynamics method \cite{Durlofsky_Brady_Bossis_1987,Sierou_Brady_2001},
where the asymptotic analytical functions for particle hydrodynamic traction multipoles are tabulated and added to a far-field expansion of the mobility matrix.
The analytic functions for multipoles are truncated at the stresslet level.
Therefore even at $\epsilon/a=0.1$, at most two digits of accuracy is achieved \cite{wilson_stokes_2013}.
Even with a low order grid $p=8$, the VSH approach has accuracy on par with Stokesian Dynamics.

For dynamic simulations with many possible collisions, to reduce
computation cost we generally use a low order grid with $p=6,8$, etc.,
because when particles are close to contact the benefits of increasing
$p$ are not significant enough to justify the extra cost.  In
particular, we set the collision radius $a_c$ to be slightly larger
than the particle radius $a$, and resolve collisions for each particle
at the collision radius.  Usually $a_c/a=1.05$.  This is a common
strategy
\cite{Brady_Bossis_1988,sierou_rheology_2002,fioreFastStokesianDynamics2019}
in particle-tracking simulations in suspension mechanics, where a soft
repulsive pairwise potential which acts at the length-scale $a_c$ is
used to prevent particle overlaps. For example, in Stokesian Dynamics
an exponentially decaying pairwise repulsive force is usually added
over the length-scale $\epsilon \approx(0.01\sim0.05)a$ to prevent the
spheres from overlapping
\cite{Brady_Bossis_1988,sierou_rheology_2002}.  In this work, we
actually choose $a_c$ based on the choice of $p$, for example,
$a_c=1.01a$ when $p=24$ and $a_c=1.05a$ when $p=6$ or $8$. 
% {\bf MJS:
%   This is messed up. Should it be $a_c=1.01a$, etc? Please correct. WY: corrected this.}

\subsection{Two spheres approaching each other}
In this section we probe the behavior of the mobility solver and the
LCP solver with a dynamic problem, where two spheres of equal radius
$a$ are dragged by constant equal and opposite external forces
$\pm\bF$ along the horizontal $\bx$-axis, with the spheres starting
with a vertical offset $a$ from each other.  The two spheres then
approach and roll over each other.  The collision radius is set to be
$a_c=1.01a$, i.e., the collision force is non-zero when the separation
$\epsilon/a=0.02$.

The orientation of the spherical harmonics grid with $p=24$ is
randomly chosen for each sphere, and different random orientations are
used for computing $\bUcal_{nc}$ and $\bUcal_c$.  The time-step 
$\Delta t=0.1\eta a^2/F$ is fixed for the results reported in
\pr{fig:pairshear}, and the simulation remains stable if $\Delta t$ is
increased by a factor of 10.  The residual tolerance of BBPGD is set
to $10^{-5}$.  The simulation shown in \pr{fig:pairshear} includes 4
stages, as illustrated by the snapshots (A), (B), (C), and (D).  In
each snapshot, the left panel shows the hydrodynamic traction induced
by the external force $\bF$, with the grey arrows showing the total
velocity $\bU_{nc}+\bU_c$ for each sphere.  The right panel shows the
hydrodynamic traction induced by collision forces.  The performance of
BBPGD and APGD solvers for the LCP are compared in
\pr{fig:pairshearBBPGD}.

In \pr{fig:pairshear}(A), the particles are close enough and are
determined to be possibly in contact.  The LCP~(\ref{lcp:LCPdef}) is
constructed as a scalar problem, because there is only one possible
contact.  The LCP solver then determines that no actual collision
happens, as shown by the zero hydrodynamic traction in the right panel
of \pr{fig:pairshear}(A).  In \pr{fig:pairshear}(B), the LCP is
constructed similarly, but the LCP solver finds that the collision
force is non-zero.  This is shown by the non-zero hydrodynamic
traction induced by collision in \pr{fig:pairshear}(B).  In
\pr{fig:pairshear}(C), the particles are about to roll over each other
so the collision force is tiny.  In \pr{fig:pairshear}(D), the
collision force is zero because the spheres are instantaneously touching each other and are moving towards a collision-free configuration. 
% {\bf MJS: Instantaneously touching each other? WY: cleared up this sentense.}

\begin{figure}[htbp]
    \centering
    \includegraphics[width=\linewidth]{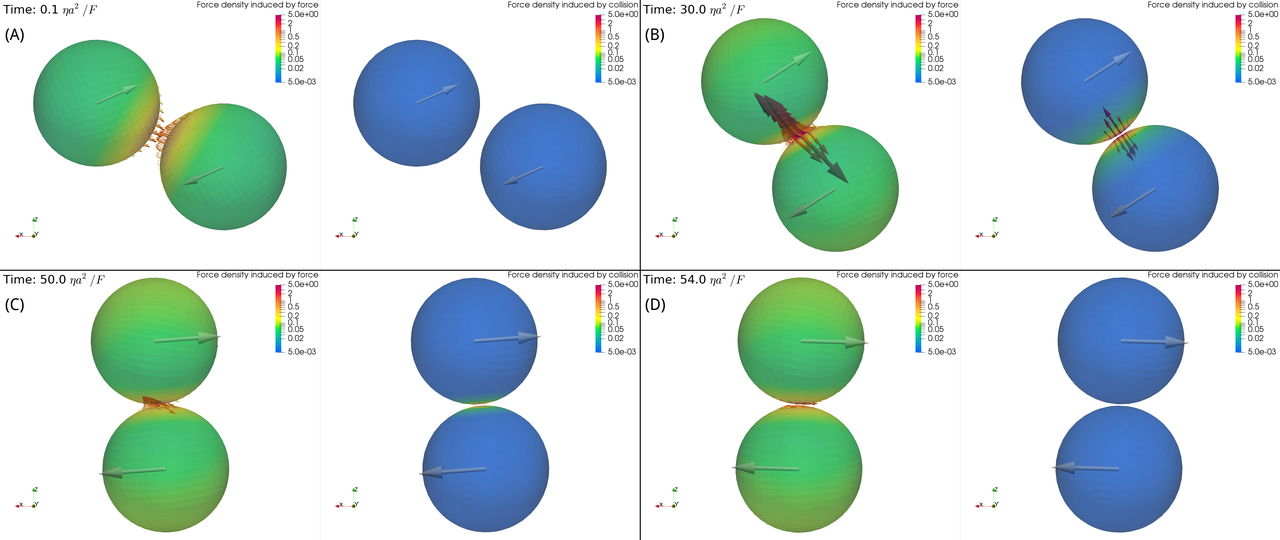}
    \caption{\label{fig:pairshear}
        A pair of particles dragged by equal and opposite external forces to roll over each other.
        In (A), (B), (C), and (D), the left panel shows the hydrodynamic traction induced by the external force $\bF$ with colored surface and vectors on the spherical harmonic grid.
        The grey arrows show the total velocity $\bU_{nc}+\bU_c$ of each particle.
        The right panels show the hydrodynamic traction induced by the collision force.
    }
\end{figure}

The performance history of BBPGD and APGD, shown in \pr{fig:pairshearBBPGD}, shows the difference between the four stages corresponding to the snapshots.
The `steps' count shows how many GD steps are used in the solver, and the `MVOPs' number shows how many evaluations of $\bM\bgamma$ are invoked during the GD steps, for each time-step.
The actual cost (and running time) scales with the number of `MVOPs' because one mobility problem is solved within each evaluation of $\bM\bgamma$.
For stages (A) and (D), only 1 MVOP is necessary for BBPGD because the solver finds that the zero initial guess of collision force already solves the problem, without the necessity of computing GD steps.
In stage (B) and (C0), BBPGD is able to converge with only 1 GD step for this scalar LCP.
Recall in \pr{alg:BBPGD} that the first BBPGD step takes 2 MVOPs, and then only 1 MVOP per step.
It is clear that in all cases BBPGD has much lower cost compared to APGD.
We observed similar performance advantages of BBPGD in many-body simulations, and all the rest results reported in this paper are computed with BBPGD only.

\begin{figure}[htbp]
    \centering
    \includegraphics[width=0.8\linewidth]{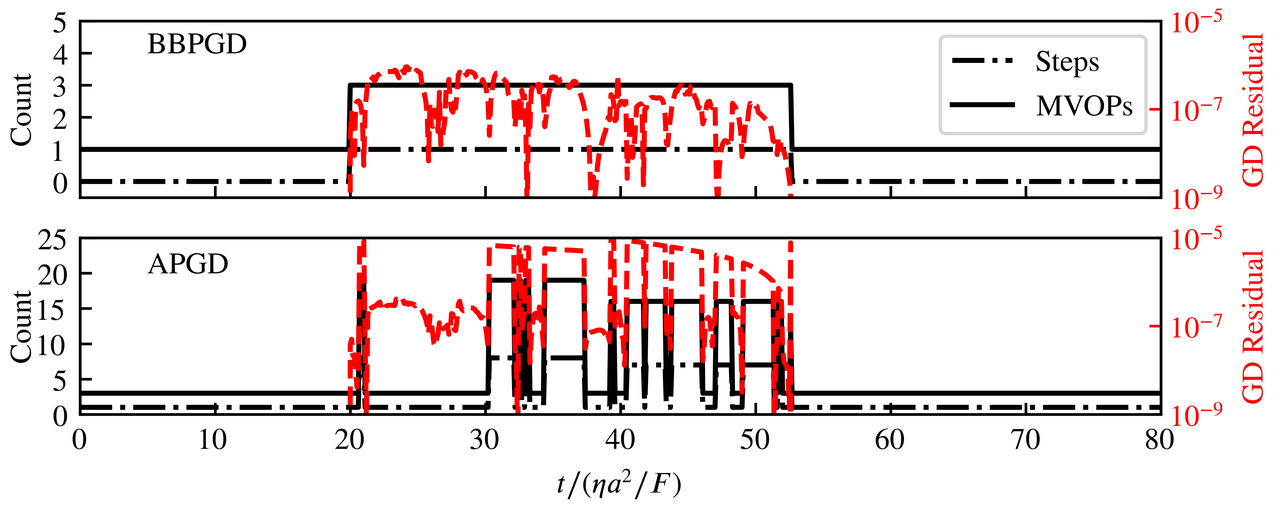}
    \caption{
        \label{fig:pairshearBBPGD}
        The performance of BBPGD and APGD solvers at each time-step for the
        simulation shown in \pr{fig:pairshear}.  The residual is set to
        $10^{-5}$ for both cases.  BBPGD significantly outperforms APGD.
    }
\end{figure}

\subsection{Sedimentation of a dense cluster}

The sedimentation of 1000 monodisperse spheres is simulated as a more demanding and realistic testing problem.
The spheres are initially placed in a dense spherical cluster of volume fraction $\approx 20\%$, and the dynamics are tracked in an unbounded fluid domain.
Each sphere is of the same radius $a$ and sediments due to the same gravitational force $\bF$ pointing towards $-z$ direction.
The collision radius for each particle is set to $5\%$ larger than $a$.
The time-step is set to $\Delta t = 0.05 \eta a^2/F$.
$p=6$ is used for the spherical harmonic representation, and $m=8$ (equivalent point density along each octree box edge) is used for KIFMM evaluations.

The snapshots of this sedimentation process are shown in \pr{fig:sedimono1k}, where for snapshots (A), (B) and (C) the four panels show the hydrodynamic traction induced by gravitational force, the hydrodynamic traction induced by collision forces in the cluster, the collision force network corresponding to the set of constraints $\bAcal$, and the velocity of each particle.
In general, particles in a crowded environment sediment faster than isolated particles, because their close neighbors drag fluid together with them, effectively reducing the relative friction between particle and fluid.
Therefore, the particles close to the cluster surface sediment much slower than the particles close to the cluster core.
These slower particles tend to lag behind and accumulate at the trailing point of the cluster.
At that point, these slow particles form a dilute structure locally, and become even slower as shown in snapshots (B) and (C) of \pr{fig:sedimono1k}.
Eventually, a tail of slow particles form behind the sedimenting cluster, as illustrated in snapshot (D) of \pr{fig:sedimono1k} showing the structure at the end of the simulation $t=200\eta a^2/F$, where each particle is colored by the hydrodynamic traction induced by collisions.
This is a well-known phenomena \cite{guazzelliSediment2006}, and has also been observed in sedimenting clouds of fibers \cite{parkCloudRigidFibres2010}.

\begin{figure}[htbp]
    \centering
    \includegraphics[width=\linewidth]{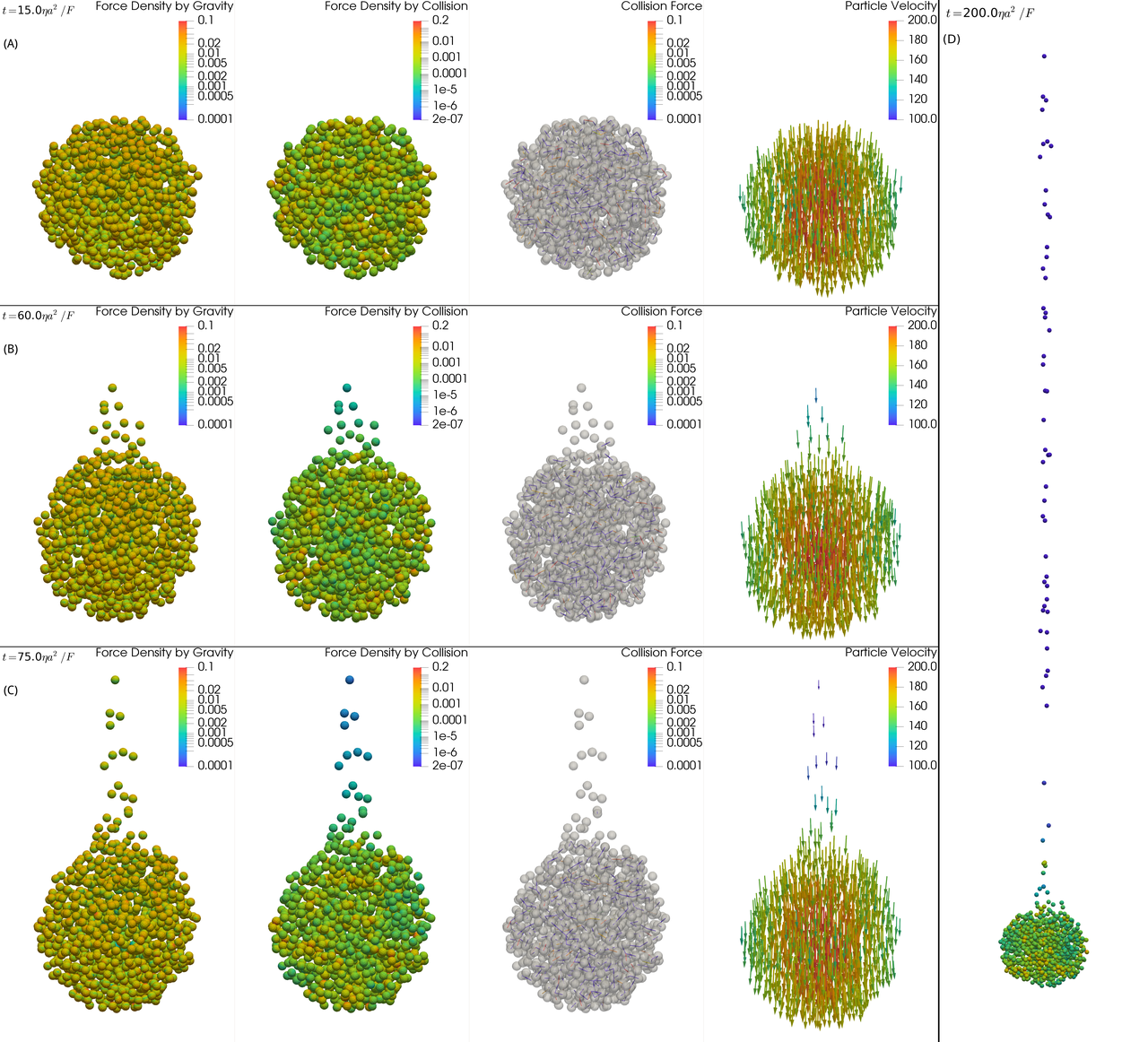}
    \caption{\label{fig:sedimono1k} The sedimentation of a dense
        cluster of $1000$ monodisperse spheres with volume fraction
        $\approx 20\%$.  In the snapshots (A), (B) and (C), the four
        panels show the hydrodynamic traction induced by the
        gravitational force $\bF$ pointing toward $-z$ direction, the
        hydrodynamic traction induced by collision forces in the
        cluster, the collision force network corresponding to the set of
        constraints $\bAcal$, and the velocity of each particle.  The
        snapshot (D) shows the structure at the end of the simulation,
        at $t=200\eta a^2/F$, where each particle is colored by the
        hydrodynamic traction induced by collisions.  A video of this
        sedimentation process is available in the Supplemental Material.
    }
\end{figure}

Because of this phenomena, the cluster of spheres becomes less dense during sedimentation.
Consequently, the number of BBPGD steps necessary to resolve collisions gradually decreases.
The performance of the BBPGD solver is shown in \pr{fig:sedimono1kBBPGD}, where the convergence tolerance of BBPGD is set to $\epsilon_{tol}=10^{-4}$.

\begin{figure}[htbp]
    \centering
    \includegraphics[width=0.8\linewidth]{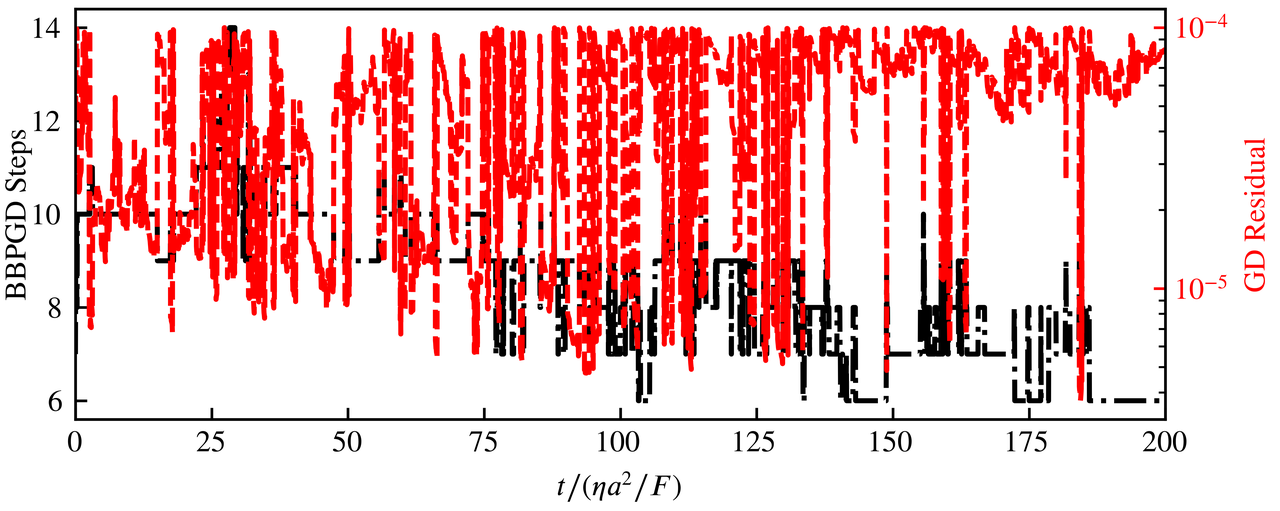}
    \caption{\label{fig:sedimono1kBBPGD} The performance of the BBPGD
    solver for the sedimentation simulation shown in
    \pr{fig:sedimono1k}.  The simulation starts from a collision-free
    configuration.  The time-step is $\Delta t = 0.05 \eta a^2/F$. The
    convergence tolerance of BBPGD is set to
    $\epsilon_{tol}=10^{-4}$.  }
\end{figure}

\subsection{Assessing the effect of numerical parameters}
In our computational framework there are several tunable parameters,
including the order $p$ of spherical harmonic representation, the
time-step $\Delta t$, the BBPGD solver tolerance $\epsilon_{tol}$, and
the collision radius $a_c$. In this section we report the effect of
these four parameters on the accuracy of the solver by repeating the
sedimentation cluster simulation of 1000 spheres done in the previous
subsection but for a short amount of time $t=10 \eta a^2/F$.  The
simulation error is measured by the relative $L_2$ error of the vector
$\bUcal\in R^{6000}$ at the end of the simulation, relative to a more
accurate reference $\bUcal_{ref}$:
\begin{align}
\epsilon_2 = \frac{\NORMtwo{\bUcal-\bUcal_{ref}}}{\NORMtwo{\bUcal_{ref}}}.
\end{align}
We fix $m=10$ (equivalent point density along each octree box edge) 
for KIFMM evaluations, and the GMRES convergence residual to $10^{-6}$.
% {\bf MJS: this is the first and only
%   appearance of KIFMM order in the whole paper. What is it? WY: fixed this}

We first vary $\Delta t$ over the range $(0.02-1.0)\eta a^2/F$, while
fixing $p=12$, $a_c=1.01a$, and $\epsilon_{tol}=10^{-5}$.  The
time-stepping as implemented in \pr{eq:time-step} has first order
accuracy, and the result shown in Table~\ref{tab:dt} 
% {\bf MJS: get rid of the page number. The table numbering is unique. WY: fixed} 
shows a consistent
behavior at the smaller time-steps. 
We note that the largest time-step, $\Delta t = 1.0\eta a^2/F$ is 20 times larger than the time-step used in the previous section, and for this large time-step each particle may move $2\sim 3$ times of particle radius during each time-step. 
Nonetheless, this simulation remains stable and still
reasonably accurate in its average error. 
% {\bf MJS: Do you approve? WY: Yes}

\begin{table}[h]
   \begin{center}
    \begin{tabular}{ c|c|c|c|c } 
     \hline
     $\Delta t / (\eta a^2/F)$ & 1.0 & 0.2 & 0.1 & 0.05 \\ 
     \hline
     $\epsilon_2$ & 0.00744 & 0.00198 & 0.000931 & 0.000372 \\ 
     \hline
    \end{tabular}
    \end{center}
   \caption{The effect of varying the time-step $\Delta t$ upon the
     relative error $\epsilon_2$. The reference case uses $\Delta
     t=0.02\eta a^2/F$, with $p=12$, $a_c=1.01a$,and
     $\epsilon_{tol}=10^{-5}$.}
    \label{tab:dt}
\end{table}

We next study the effect of varying the BBPGD tolerance $\epsilon_{tol}$
between $10^{-3}$ to $10^{-6}$, while fixing other parameters to
$p=12$, $a_c=1.01a$, and $\Delta t =0.1\eta a^2/F$.  The results,
reported in Table~\ref{tab:bbpgd} 
% {\bf MJS: Get rid of the page number. WY: fixed }
, show
that tuning $\epsilon_{tol}$ has little effect on the simulation
accuracy.  This is because the most difficult part of the BBPGD solver
is the identification of the active set $\bAcal_c$, i.e., determining
where are the actual collisions.  
Once this set is identified, the
BBPGD solver converges quickly, as demonstrated by the average number
of BBPGD steps in Table~\ref{tab:bbpgd}, where decreasing $\epsilon_{tol}$ by
an order of magnitude only requires 2 or 3 more BBPGD steps. \citet{Dai_Fletcher_2005} have thoroughly analyzed this behavior of BBPGD algorithm. 
% {\bf
%   MJS: What is the point of this remark? Does it mean explained by
%   them, or also observed by them? WY: rephrased this description.}.
\begin{table}[h]
    \begin{center}
     \begin{tabular}{ c|c|c|c } 
      \hline
      $\epsilon_{tol}$ & $10^{-3}$ & $10^{-4}$ & $10^{-5}$  \\ 
      \hline
      $\epsilon_2$ & $2.13\times10^{-7} $ & $3.52\times10^{-8} $ & 0 \\ 
      Average BBPGD Steps & 9 & 12 & 14 \\
      \hline
     \end{tabular}
     \end{center}
    \caption{The effect of varying the tolerance $\epsilon_{tol}$ of
      BBPGD on the relative error $\epsilon_2$. The reference case
      uses $\epsilon_{tol}=10^{-6}$. Other parameters are fixed to
      $p=12$, $a_c=1.01a$, and $\Delta t =0.1\eta a^2/F$. 
      $\epsilon_2=0$
      means the error has fallen below machine precision. 
    %   {\bf MJS: Only zero is below machine precision... :) WY: fixed it}
      }
    \label{tab:bbpgd}
\end{table}

The effect of changing the order of the spherical harmonics expansion
is shown in Table~\ref{tab:psh} 
% {\bf MJS: Get rid of page numbers WY: fixed}
, where $p$
is varied from 8 to 20, with $p=20$ used as the reference case, and
with other parameters held fixed at $a_c=1.01a$,
$\epsilon_{tol}=10^{-5}$, and $\Delta t =0.1\eta a^2/F$. The error
$\epsilon_2$ shows slow convergence with increasing $p$. This is
because particles are very close to each other in this dense cluster
simulation, and as we have shown in \pr{fig:lubtriangle}, the error of
the mobility solver decreases slowly in this regime.  This is the
intrinsic difficulty of resolving lubrication effects efficiently in
dynamic simulations, and remains an open problem.  The reference data
used in \pr{fig:lubtriangle} is computed by a highly accurate method
which is not generally feasible in dynamic simulations. Because of the
cost in increasing the order of the spherical harmonics expansion, we
do not use high order expansions in the large scale simulations of the
next section .
\begin{table}[h]
    \begin{center}
     \begin{tabular}{ c|c|c|c } 
      \hline
      $p$ & 8 & 12 & 16  \\ 
      \hline
      $\epsilon_2$ & 0.00697 & 0.00268 & 0.00114 \\ 
      \hline
     \end{tabular}
     \end{center}
    \caption{The effect of varying $p$, the order of the spherical
      harmonics expansion, upon the relative error
      $\epsilon_2$. The reference case uses $p=20$, and other
      parameters are fixed to $a_c=1.01a$, $\epsilon_{tol}=10^{-5}$,
      $\Delta t =0.1\eta a^2/F$.}
    \label{tab:psh}
\end{table}

Finally, we investigate the effect of the collision radius $a_c$.
Table~\ref{tab:ac} 
% {\bf MJS: Get rid of the page number. WY: fixed} 
reports the
relative error $\epsilon_2$ for $a_c$ varying from $1.001a$ to
$1.05a$, and fixing $p=12$, $\epsilon_{tol}=10^{-5}$, and $\Delta t
=0.1\eta a^2/F$. $\epsilon_2$ decreases monotonically with decreasing
$a_c$, but the convergence is slow. This is again related to the
difficulty in resolving lubrication effects when particles are close
to each other, because changing $a_c$ only affects those particles
that may get closer.  In this regime the accuracy of spherical
harmonics expansion is only modestly accurate, even with a high order
expansion. Therefore, there is generally no need to pick a very small
value of $a_c$, and we will use fairly large values of $a_c$ in the next
section for large scale simulations.
\begin{table}[h]
    \begin{center}
     \begin{tabular}{ c|c|c|c|c } 
      \hline
      $a_c/a - 1$ & 0.05 & 0.02 & 0.01 & 0.005  \\ 
      \hline
      $\epsilon_2$ & 0.0108 & 0.00656 & 0.00521 & 0.00352 \\ 
      \hline
     \end{tabular}
     \end{center}
    \caption{The effect of varying the collision radius $a_c$ upon the
      relative error $\epsilon_2$. The reference case uses
      $a_c=1.001a$, with other parameters fixed to $p=12$,
      $\epsilon_{tol}=10^{-5}$, and $\Delta t =0.1\eta a^2/F$.}
    \label{tab:ac}
\end{table}

\subsection{Scaling benchmark}
The sedimentation of particles randomly placed in a cubic box in an
unbounded fluid is also used to benchmark the scaling of our
implementation, using up to $80000$ spheres on 1792 cores (64 nodes).
Weak and strong scaling is measured on a cluster where each node is
equipped with two 14-core Intel Xeon CPU E5-2680 v4 running at
2.40GHz.  Hyper-Threading is turned off.  One \texttt{MPI} rank is
launched for each CPU socket, and every \texttt{MPI} rank launches 14
\texttt{OpenMP} threads.  In total, 28 cores are used on each node,
and the nodes are interconnected by a 100 Gb/s Intel Omni-Path (OPA)
network.  Intel MPI compilers and libraries are used, together with
Intel MKL libraries for BLAS, LAPACK and FFTW3 functions.  The radius
$a$ of each particle is randomly generated from a log-normal
distribution with standard parameters $\mu=a$ and $\sigma=0.3a$.  The
collision radius $a_c=1.05a$ is taken for each sphere.  The time-step
is fixed at $\Delta t=0.1 \eta a^2/F$, where $\bF$ is the
sedimentation force applied on each particle.  $p=8$ is used for the
spherical harmonic grid and $m=10$ (equivalent point density along
each octree box edge) is used for the KIFMM evaluations, throughout
these benchmarks.  The largest system of $8\times10^4$ spheres has
$486$ degrees of freedom for hydrodynamic traction on each sphere and
$\sim\sctf{3.9}{7}$ degrees of freedom in total.

\begin{figure}[htbp]
    \centering
    \includegraphics[width=0.8\linewidth]{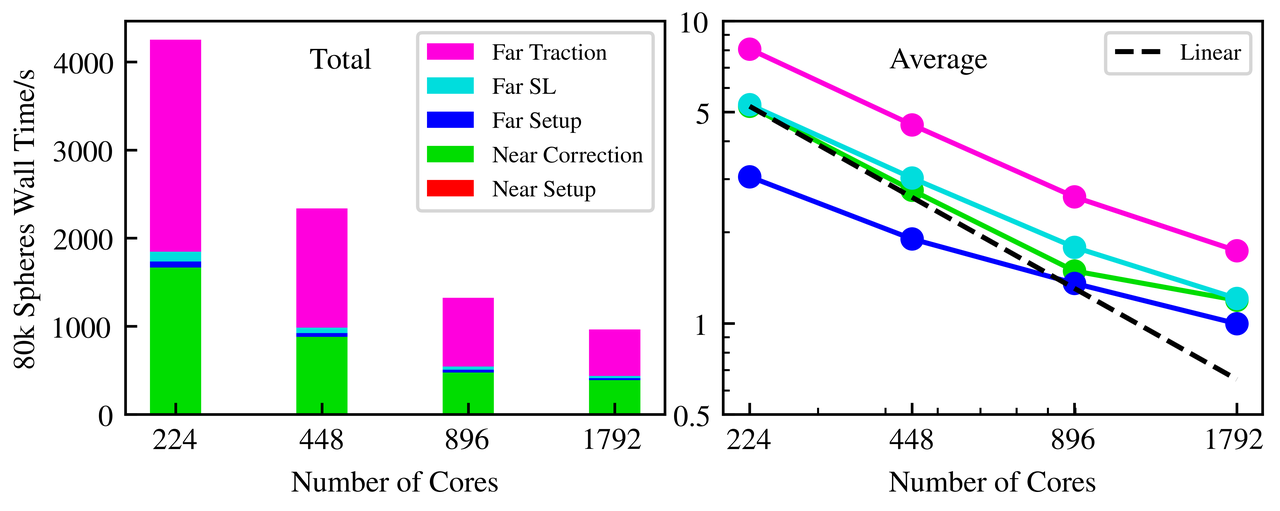}
    \caption{
        \label{fig:strong80k}
        The strong scaling of $8\times 10^4$ spheres for 11 time-steps
        on up to 1792 cores.  The left panel shows the total time for
        each operation.  `Far Setup' and `Near Setup' both run 22
        times.  `Far SL' runs 21 times and `Far Traction' runs 298
        times.  `Near Correction' runs $21+298=319$ times for both SL
        and Traction corrections.  The number of times each operation
        is executed coincides with the previous smaller scale test of
        $10,000$ spheres because the configuration is generated at
        similar radius distribution and volume fraction.  The right
        panel shows the average time for each operation.  }
\end{figure}

\begin{figure}[htbp]
    \centering
    \includegraphics[width=0.8\linewidth]{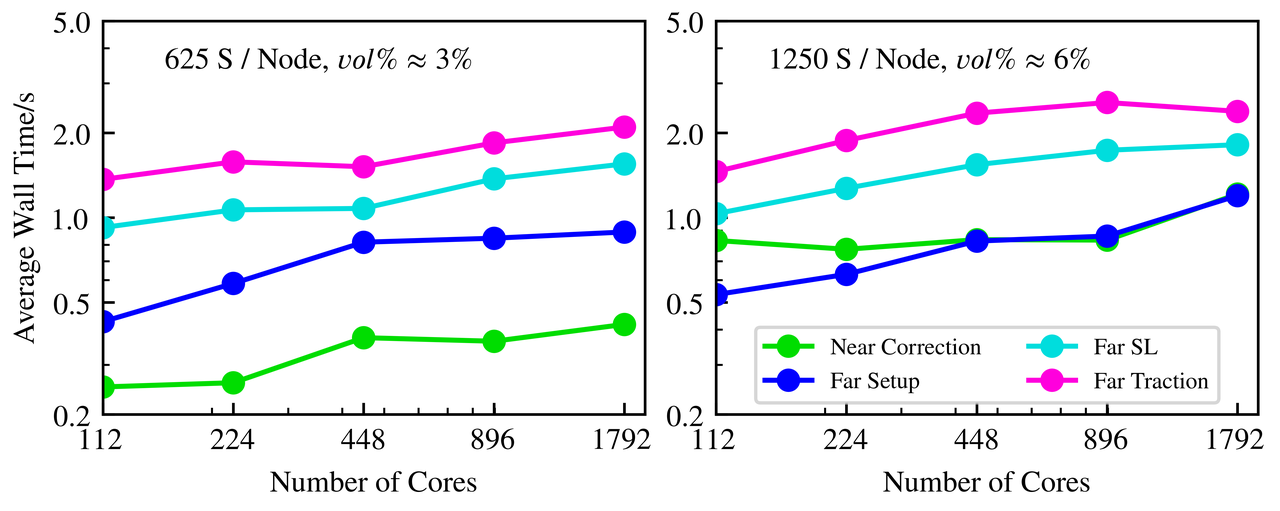}
    \caption{
        \label{fig:weak}
        The weak scaling at different volume fractions. The time for
        performing each operation once is measured.  The left panel
        shows the scaling with 625 spheres per node, ranging from 2500
        spheres (4 nodes) to $\sctf{4}{4}$ spheres (64 nodes).  The
        right panel shows the scaling with spheres per node, ranging
        from $5000$ spheres (4 nodes) to $80,000$ spheres (64 nodes).
    }
\end{figure}

The running time for the five major operations are measured and
reported in \pr{fig:strong10k}, \ref{fig:strong80k}, and
\ref{fig:weak}.  `Far Setup' refers to the time to compute the
coordinates of spherical harmonic grid points and to build an adaptive
octree for KIFMM evaluations.  `Far Traction' refers to the evaluation
of the traction operator $\STr$ in the BI equation
\pr{eq:rigidbodyrho}.  `Far SL' refers to the evaluation of the single
layer operator $\SSg$ to evaluate the fluid velocity $\bu$ on grid
points after the traction $\rho$ is found by solving
\pr{eq:rigidbodyrho}.  `Near Setup' refers to the time to perform near
neighbor detection to prepare the VSH evaluations.  `Near Correction'
refers to the time for VSH corrections.  The cost of constructing the
LCP sparse matrix $\bDcal$ and applying the sparse matrix-vector
multiplications during LCP solution is not shown here because it is
negligible compared to the total cost of mobility solutions.

\pr{fig:strong10k} shows the strong scaling for $10,000$ spheres on up
to 224 cores (8 nodes) and \pr{fig:strong80k} shows the strong scaling
for $80,000$ spheres on up to 1792 cores (64 nodes).  In both cases
the same initial configuration at approximately $6\%$ volume fraction
is used throughout all runs on different numbers of cores.  The left
panels of \pr{fig:strong10k} and \ref{fig:strong80k} show the total
wall time for different operations for 11 time-steps.  The right panels
show the average wall time for performing each operation once.  The
black dashed line denotes the ideal scaling with $100\%$ parallel
efficiency for reference.  In these scaling tests, spherical harmonic
grids with the same order $p$ but different orientation are used for
mobility and collision problems.  Therefore, at each time-step `Far
Setup' and `Near Setup' are both executed twice, i.e., once for the
mobility problem and once for the collision problem.  The `Far
Traction' is evaluated once at each GMRES iteration of solving the
mobility problem \pr{eq:rigidbodyrho}, but the `Far SL' is evaluated
only when GMRES converges.  In general, `Far Setup' and `Far SL' are
executed the same number of times.

\subsubsection{Strong scaling}
\begin{figure}[htbp]
    \centering
    \includegraphics[width=0.8\linewidth]{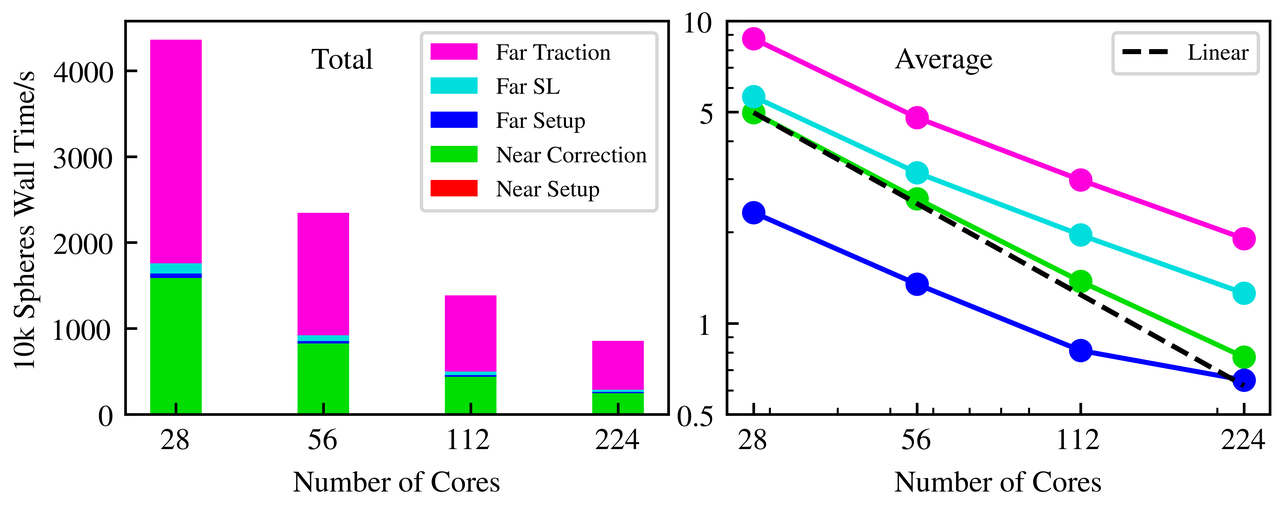}
    \caption{
        \label{fig:strong10k}
        The strong scaling of $10,000$ spheres for 11 time-steps on up
        to 224 cores.  The left panel shows the total time for each
        operation.  `Far Setup' and `Near Setup' both run 22 times.
        `Far SL' runs 21 times and `Far Traction' runs 298 times.
        `Near Correction' runs $21+298=319$ times for both SL and
        Traction corrections.  The right panel shows the average time
        for each operation.  }
\end{figure}

The parallel efficiency of the `Near Correction' part is close to ideal except for the largest case on 1792 cores, shown in \pr{fig:strong80k}, where we suspect there is some load-unbalancing due to domain decomposition.
The efficiency of `Far Traction' and `Far SL' are a bit lower than the near correction part, but they are consistent with the results demonstrated for the library PVFMM \cite{malhotra_pvfmm_2015}, which we used to perform KIFMM evaluations.
The efficiency of `Far Setup' is even lower, but since it runs only once per time-step and takes only a small fraction of the total running time, there is almost no noticeable benefit in optimizing this part.
The wall time for `Near Setup' is negligible and invisible in both \pr{fig:strong10k} and \pr{fig:weak}.

\subsubsection{Weak scaling}
Since the cost of performing near corrections grows with the number of near pairs, we keep the system volume fraction approximately constant by adjusting the box size according to the number of particles.
\pr{fig:weak} shows the average time for performing each operation once during a 10-time-step simulation, for two volume fractions $3\%$ and $6\%$.
Ideally a flat line is expected if the parallel efficiency is 100\%.
Here the running time of `Far Setup' grows faster but other operations are not significantly far from the ideal case.
This non-ideal scaling of `Far Setup' does not matter in real simulations because this operation is performed only once when particles move, that is, once for each time-step.
Therefore the net cost of this `Far Setup' is far less than the total cost in other operations, as can be seen from the left panels in \pr{fig:strong10k} and \pr{fig:strong80k}.

\section{An active-matter case study: Suspensions of Stokes rotors}
\label{sec:application}

Recently, suspensions of active particles have been intensely
investigated as real-world realizations of active matter
\cite{SS2013}.  Active matter refers to multiscale materials whose
microscopic constituents perform directed work on the system, leading
to large-scale dynamics such as self-organization.  A canonical
example of an active suspension is a bacterial bath, wherein
microorganisms interact through the flow-fields created by their
self-propulsion \cite{DCCGK2004}.  These systems can evince
instabilities towards alignment and unpredictable large-scale flows
sometimes termed ``bacterial turbulence'' \cite{SS2008a}. Similarly
complex dynamics appears in suspensions of microtubules which are
``polarity sorted'' by the directed motion of cross-linking
motor-proteins \cite{SanchezEtAl2012,Shelley2016,ND2017}.

A very different kind of active suspension consists of immersed
particles that are driven to rotate, say rather than swim or sort,
with that rotation again creating flow fields that can create
large-scale coupling and dynamics.  Such rotor systems are typically
-- but not always \cite{kuimovaMappingViscosityCells2012,OSS2019} --
driven by external means, such as a rotating magnetic field. Such
systems can show activity-induced phase separation
\cite{yeoCollectiveDynamicsBinary2015}, crystallization
\cite{OSS2019}, odd surface dynamics and rheology \cite{SoniEtAl2019},
and forms of active turbulence \cite{kokotActiveTurbulenceGas2017}.

Here we use the methods developed here to study the dynamics of
closely packed rotor systems, showing the development of large-scale
dynamics.  In each example we assume the same external torque $\bT$ is
applied to every particle.  Within each time-step, we first solve a
mobility problem to compute the velocity $\bUcal_{nc}$ driven by
$\bT$.  This is followed by applying the collision resolution method
described in \pr{sec:formulation} to compute the collision velocity
$\bUcal_c$.  A fixed time-step $\Delta t$ is used throughout, and taken
large enough so that each particle may move about $20\%$ of its radius
at each time-step.  This time-step is roughly two to three orders of
magnitude larger than the usual choices in Stokes suspension
simulations \cite{Foss_Brady_2000,Sierou_Brady_2001}.  We do this to
demonstrate the stability of our collision resolution algorithm.  For
all simulations in this section, $p=6$ is used for the spherical
harmonic grid and $m=8$ (equivalent point density along each octree
box edge) is used for KIFMM evaluations.

\subsection{A cluster of $10,000$ rotors}
We first probe the dynamics of a spherical cluster of $10,000$
polydisperse (in diameter) spherical rotors in an unbounded fluid.  A
constant torque $\bT$ along the $z$-axis is applied to each sphere.
The radius of each particle is randomly generated from a log-normal
distribution with standard parameters $\mu=a$ and $\sigma=0.3a$, where the probability density function is defined as $p(x)=\frac{1}{x \sigma \sqrt{2 \pi}} \exp \left(-\frac{(\ln x-\mu)^{2}}{2 \sigma^{2}}\right)$.  The
cluster is approximately spherical and the volume fraction of spheres
is approximately $10\%$.  A fixed time-step, $\Delta t = 1.0 \eta a^3
    /T$, is used.  The collision radius $a_c$ is set to $10\%$ larger than
the radius for each particle.

\begin{figure}[htbp]
    \centering
    \includegraphics[width=\linewidth]{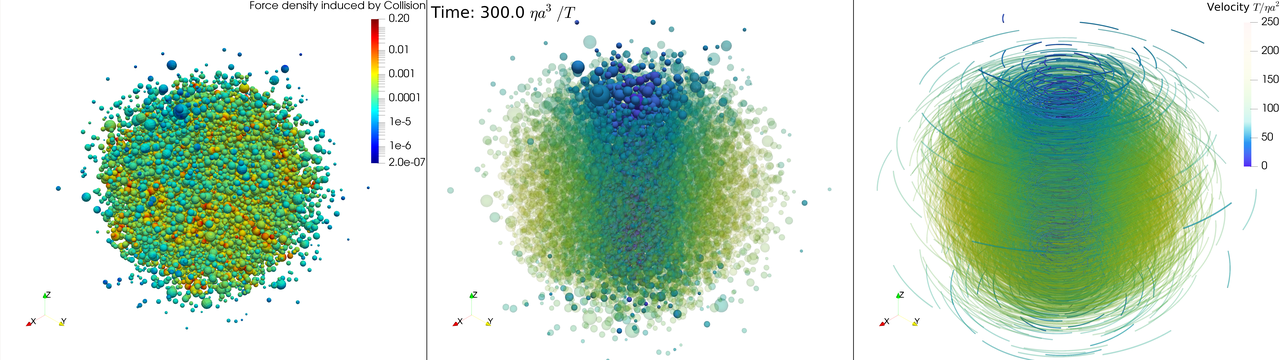}
    \caption{
        \label{fig:Fig_RotorPoly10k}
        Snapshots at $t=300 \eta a^3/T$ of a cluster driven by a fixed
        torque $\bT$ on each sphere. The left panel shows the
        hydrodynamic traction magnitude across each rotor surface.
        The middle and the right panels are both colored by the
        instantaneous center velocity magnitude of each
        sphere at $t=300 \eta a^3/T$, showing the particle structure
        at this time-step and the particle trajectories starting from
        the initial configuration, respectively.  A video of this
        simulation is available in the Supplemental Material.  }
\end{figure}

In the absence of hydrodynamic interactions, each particle would
rotate at constant rate about its $z-$axis, with larger particles
rotating more slowly than smaller ones (given the constant driving
torque).  Figure \ref{fig:Fig_RotorPoly10k}, at $t=300\eta a^3/T$,
shows the effect of hydrodynamic and steric coupling in creating
large-scale rotation of the cluster.  The left panel shows the
hydrodynamic traction magnitude across each rotor surface.  The middle
panel shows the instantaneous velocity magnitude of each particle,
where blue ones are slow and green ones are fast.  The right panel
shows the trajectory of each particle, starting from $t=0$, and
colored by each particle's velocity magnitude at each time-step.  The
trajectories show that this cluster is rotating relative to a common
axis in the $z$ direction through the cluster center.  The simulation
runs till $t=500\eta a^3/T$, and no visible expansion or shrinking of
this cluster is observed.

To analyze this global rotation, we set up a cylindrical coordinate
system $(r,\theta,z)$, where $r,z=0$ is fixed at the geometric center
of the cluster.  
We project the velocity $\bU$ of each particle onto
this cylindrical coordinate system and take the angular component
$U_\theta$.  Then we estimate the normalized distribution
$P(U_\theta,r)$ using data accumulated over this entire simulation.
The result is shown as a two dimensional histogram in
\pr{fig:Fig_RotorPoly10kRVel}.  The distribution $P(U_\theta,r)$ shows
that the overall motion is close to that of a rigid body rotation
$U_\theta \propto r$, where particles within the cluster all rotate
about the central $z$-axis with roughly the same angular velocity.
Only relatively few particles, far from the cluster center with
approximately $r/a>45$, moves more slowly than the cluster's global
rotation.  The expectation is that collectively induced velocities
will decay as $r^{-2}$ for $r>>1$, as the cluster will appear as a
rotlet singularity in the far-field.

\begin{figure}[htbp]
    \centering
    \includegraphics[width=0.8\linewidth]{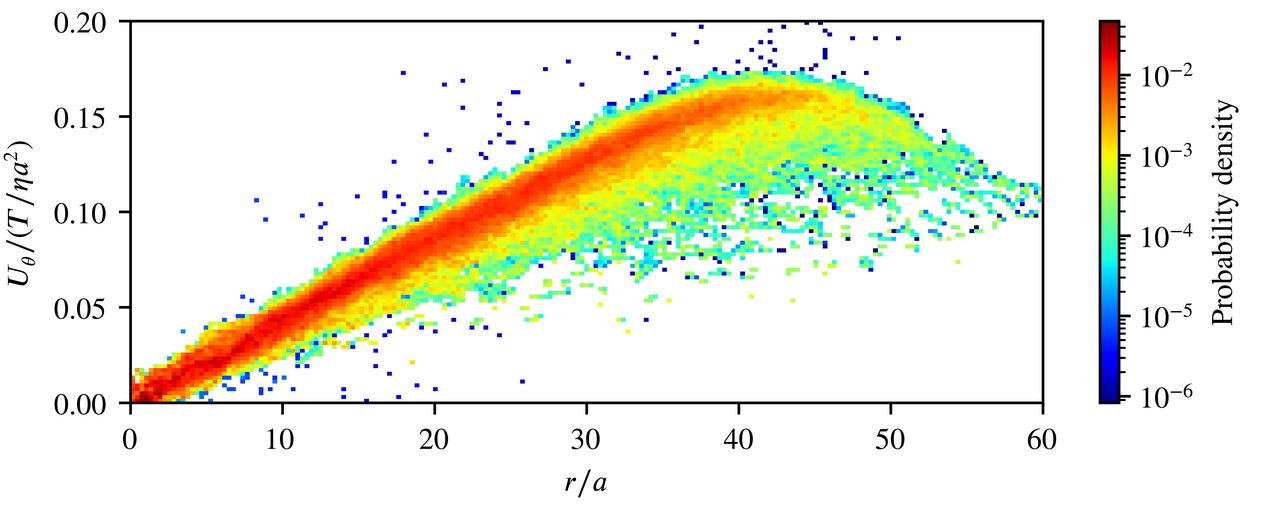}
    \caption{
        \label{fig:Fig_RotorPoly10kRVel}
        The normalized velocity distribution $P(U_{\theta},r)$ for
        rotors in the cluster. The distribution $P$ is normalized so
        that $\int_0^\infty\int_0^\infty P(U_\theta,r) 2\pi r drdU=1$.
        The data is accumulated over the entire simulation of $t=500\eta
            a^3/T$.  }
\end{figure}

In this system, for each time step approximately $2000$ possible
collisions are included in the collision resolution solver, and
approximately $250$ collisions actually happen.  That is, the
size of the set $\bAcal$ is roughly $2000$ and the size of the set
$\bAcal_c$ is roughly $250$.  The performance of the collision
resolution algorithm over the course of the simulation is shown in
\pr{fig:rotorpoly10kBBPGD}.  The simulation starts from a
collision-free configuration, and then collisions are generated and
reaches a steady state as the cluster remains close to a rigid body
rotation, as shown above.  In the beginning roughly 14 BBPGD steps are
necessary to resolve the collisions and later this number increases to
roughly 20.  In comparison to the sedimentation case reported in
\pr{sec:results}, this rotor simulation involves 10 times the number
of particles but the number of necessary BBPGD steps only slightly
increases.  Empirically, the number of BBPGD steps scales much slower
than the number of particles.  This feature makes this collision
resolution algorithm suitable for large scale simulations.

\begin{figure}[htbp]
    \centering
    \includegraphics[width=0.8\linewidth]{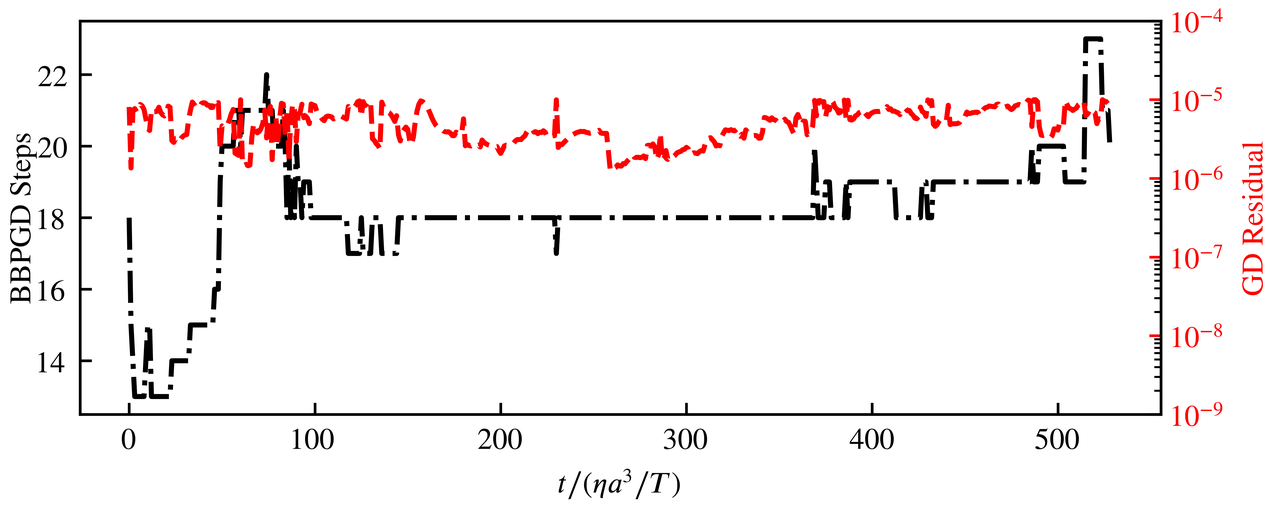}
    \caption{
    \label{fig:rotorpoly10kBBPGD}
    The performance of the BBPGD solver for $10,000$ rotors in a
    spherical cluster.  The simulation starts from a configuration
    with many collisions.  The time-step $\Delta t = 1.0 \eta
        a^3/T$.  The convergence tolerance of BBPGD is set to
    $\epsilon_{tol}=10^{-5}$.  }
\end{figure}

\subsection{The dynamics within a monolayer of $20,000$ rotors}

Recent experiments and simulations have studied the dynamics of rotors
confined to a surface within a fluid
\cite{yeoCollectiveDynamicsBinary2015}, or atop a solid substrate
\cite{SoniEtAl2019}, with the driving external torque perpendicular to
the surface or substrate.  We use this setup to analyze the internal
processes of a cluster of rotors, most especially the evolution of the
collision network, using a disk of $20,000$ monodisperse rotors
confined in a monolayer, as shown in \pr{fig:intro}.  We set $\Delta
t= 0.5 \eta a^3/T$.  The area fraction of particles is approximately
$60\%$ for this simulation, much denser than the previous example.  As
the collision radius of each particle is also set to $a_c=1.1a$, the
effective area fraction for collision resolution is around $70\%$.  As
a result, the rotors show a good deal of hexagonal ordering, as shown
in \pr{fig:perprotordisksnap}.  Even at such high densities, the BBPGD
collision resolution solver takes about 20 descent steps per time
step, similar to the spherical cluster example shown in
\pr{fig:rotorpoly10kBBPGD}.

\begin{figure}[htbp]
    \centering
    \includegraphics[width=\linewidth]{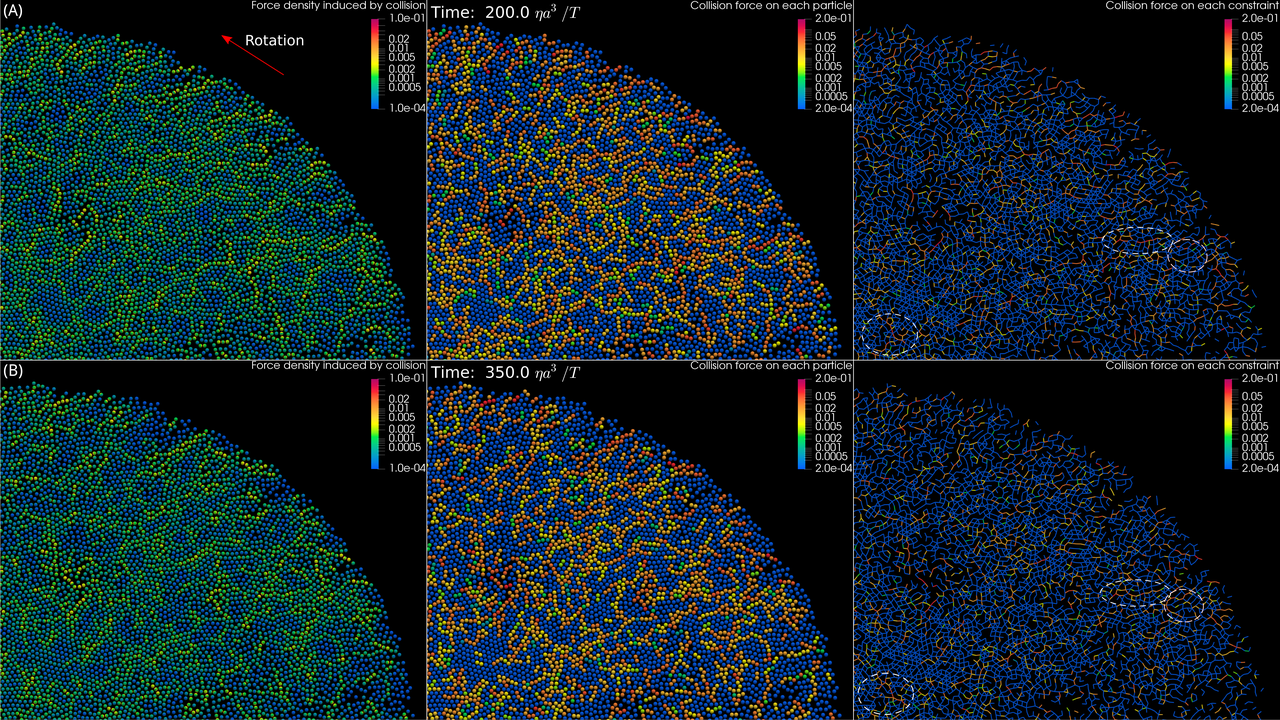}
    \caption{
        \label{fig:perprotordisksnap}
        Snapshots of a quarter of the rotor disk at times $t=200\eta
            a^3/T$ (A) and $350\eta a^3/T$ (B).  Counterclockwise global
        rotation is driven by the torque $\bT$ on each rotor,
        perpendicular to the disk.  For (A) and (B), the left panels
        show the magnitude of hydrodynamic traction induced by
        collisions on each rotor surface, the middle panel shows the
        net collision force on each particle, and the right panel
        shows the collision force $\gamma_\ell$ on each contact
        constraint $\ell$.  The rightmost dashed circle marks a set of
        rotors that form a transient collision chain appearing in (B)
        only.  Each of the other two white dashed circles on the left
        in (A) and (B) mark a single persistent collision chain.  A
        video of this simulation is available in the Supplemental
        Material.  }
\end{figure}

The dynamics of the rotors is detailed in \pr{fig:perprotordisksnap},
where the hydrodynamic force density, i.e. the traction $\bff$, the
net collision force $\bFcal_c$ on each particle, and the collision
force magnitude $\gamma$ for each contact constraint are shown in the
left, center, and right panels, respectively.  Figure
\ref{fig:perprotordisksnap} (A) and (B) show a snapshot at times
$t=200 \eta a^3/T$ and $t=350\eta a^3/T$, respectively. The
comparison between (A) and (B) clearly indicates that there are many
small-scale motions caused by collisions, while the entire cluster
is rotating collectively and differentially.

To examine this rotation, we compute the normalized distribution
$P(U_\theta,r)$ as in \pr{fig:Fig_RotorPoly10kRVel}.  This is shown in
\pr{fig:perprotordiskrvel}. The radius of the disk is $R=195a$, and
$U_\theta\propto r$ holds only near the center of the disk, and at the
disk edge we observe a rapid increase in angular velocity. To
understand this, we constructed a simple continuum model of the rotor
assembly as an infinite number of rotlets at uniform density. This
continuum model gives, for example, an angular velocity $U_\theta(r)$
with a logarithmic divergence at the disk edge:
\begin{align}
    \frac{U_\theta}{T/(\eta a^2)} \approx A\log
    \left(R-r\right)+B,\mbox{  as  }\quad r\to R.
\end{align}
This calculation is detailed in \pr{appsec:rotletdisk} for the
reader's interest. In \pr{fig:perprotordiskrvel} we show a fit (black
dashed curve) to this form, i.e. for $A$ and $B$, from our numerical
data in the range $r>170a$. It appears that the simulation has
achieved sufficient scale to describe well a continuum of forced
particles.

\begin{figure}[htbp]
    \centering
    \includegraphics[width=0.8\linewidth]{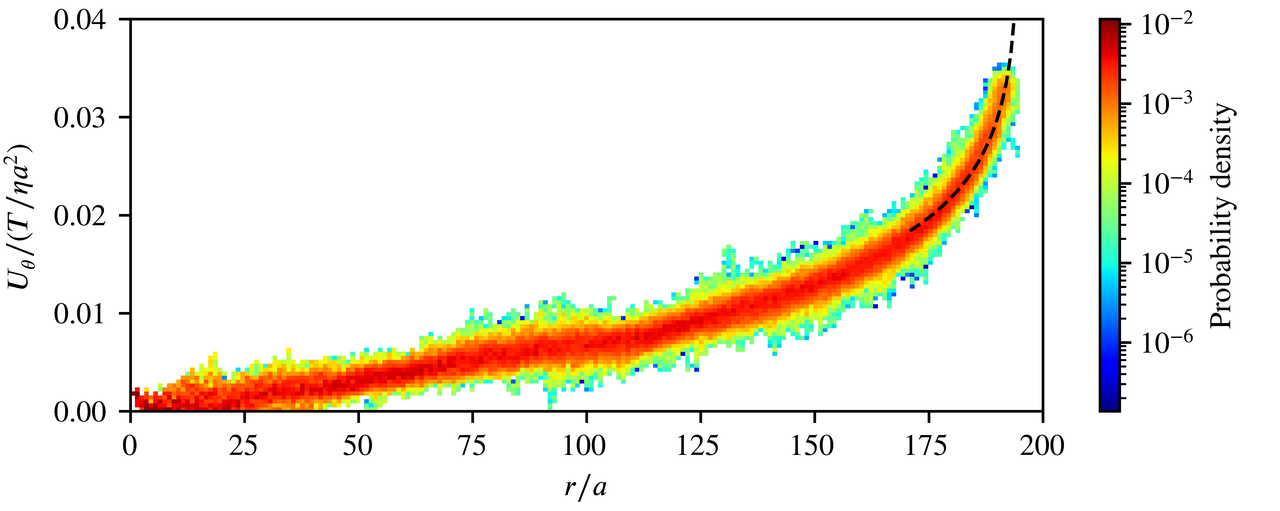}
    \caption{
        \label{fig:perprotordiskrvel}
        The normalized velocity distribution $P(U_{\theta},r)$ for rotors
        in the monolayer disk.  The distribution $P$ is normalized so that
        $\int_0^\infty\int_0^\infty P(U_\theta,r) 2\pi r drdU=1$.  The
        data is accumulated over the entire simulation of $t=350\eta
            a^3/T$.  The dashed curve is the function $A\log
            \left(R-r\right)+B$, where $A$ and $B$ are estimated by fitting to
        the simulation data in the range $r/a>170$.}
\end{figure}

To quantify the time-scales induced by collisions, we analyze the
lifetime of each collision constraint with collision force
$\gamma_\ell$, for both the set $\Acal$ for all constraints, and its
subset $\Acal_c$ for active constraints.  For the set $\Acal$ the
lifetime of a constraint is defined as the number of time-steps $k$
during which the constraint remain included in the LCP solver.  For
the subset $\Acal_c$, the lifetime is defined as the number of
time-steps during which the solution $\gamma_\ell>0$ for this
constraint $\ell$.  The normalized distributions for these two cases
are shown in \pr{fig:perprotordiskconstraintlife}.

\begin{figure}[htbp]
    \centering
    \includegraphics[width=0.8\linewidth]{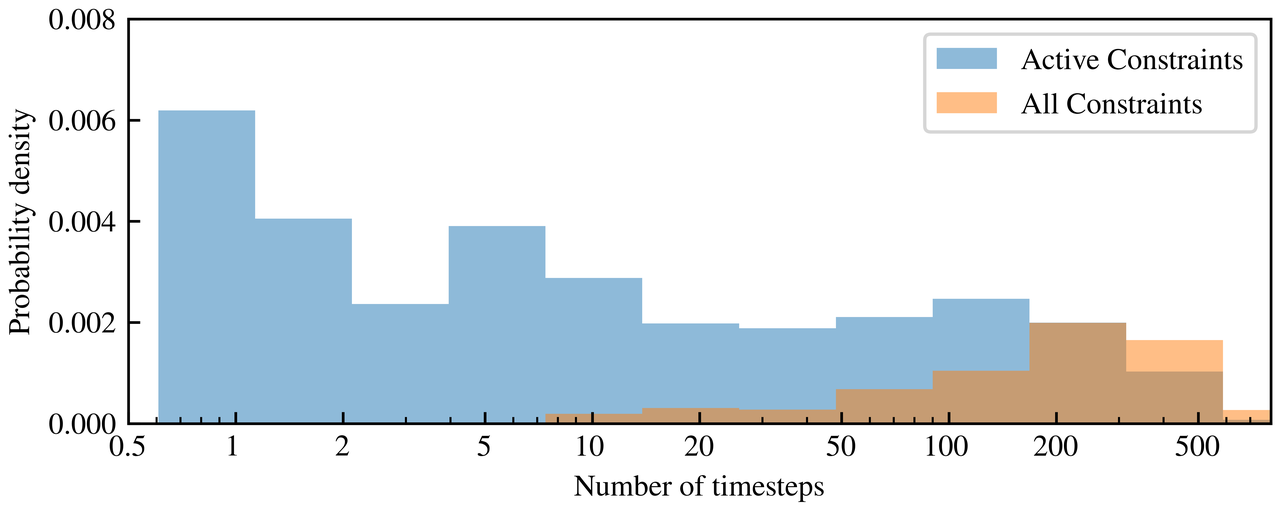}
    \caption{
        \label{fig:perprotordiskconstraintlife}
        The normalized distribution $P(k)$ of constraint lifetimes.
        $P(k)$ is normalized so that $\sum_0^{k_{\max}} P(k) = 1$, where
        $k$ is the number of time-steps, and $k_{\max}$ is the total
        number of time-steps.  The blue and yellow distributions show
        $P(k)$ for the active constraint set $\Acal_c$ and the all
        constraint set $\Acal$, respectively.}
\end{figure}

Figure \ref{fig:perprotordiskconstraintlife} shows that while many
constraints in the active set $\Acal_c$ last for only 1 time-step,
there is a broad region of short-lived constraints lasting between 2 and 20
time-steps.  These longer-lived active constraints are of relatively
lower probability, but are not rare.  However, almost no constraints
in the set $\Acal$ have a lifetime shorter than 10 time-steps.  The
majority of them last more than 200 time-steps.  This is because the
set $\Acal$ depends solely on the particles' geometric configuration.
While the rotation rate is differential in this simulation,
neighboring pairs of particles rarely change their relative positions,
and so the set $\Acal$ barely changes from step to step.  This is
shown in the right panels of \pr{fig:perprotordisksnap}(A) and (B).
In contrast, the active constraint set $\Acal_c$ depends not only on
geometry, but also on the velocity $\bUcal_{nc}$, i.e., the particles'
tendency to collide with each other.  For this many-body problem,
slight relative displacements of particles may induce significant
changes in relative velocities, and generate very different $\Acal_c$
at different time-steps.  As a final note, collisions may form
`collision force chains' when a few close-by particles run into one
another like a chain.  The two white dashed circles on the left in
\pr{fig:perprotordisksnap}(A) and (B) mark persisting collision
chains, while the rightmost white dashed circles mark a transient
collision chain appearing in (B) only.

\subsection{Dynamics of a tangentially forced layer of $20,000$ rotors}
In a last example, we place a square sheet of $20,000$ rotors on the
$z=0$ plane, and again apply a torque $\bT$ on each particle but now
along the $y$-axis.  Having the putative rotation axis aligned with
the layer is a very different kind of forcing from the previous
examples, and is conceptually akin to a vortex layer or sheet immersed
in an inviscid fluid.  We set $\Delta t=\eta a^3/T$.  The initial area
fraction is still $\approx 60\%$ and we set the collision radius $a_c
    = 1.1 a$ as in the previous examples.

\begin{figure}[htbp]
    \centering
    \includegraphics[width=\linewidth]{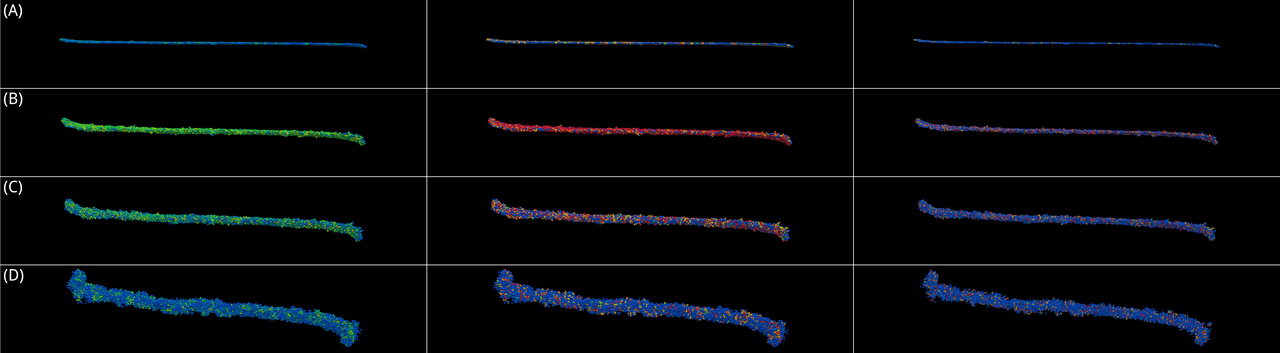}
    \caption{
        \label{fig:insrotorsquareside}
        A side view of simulation snapshots for $20,000$ rotors on the
        $z=0$ plane driven by torque $\bT$ aligned with the $y$-axis.
        The three panels of (A), (B), (C), and (D) show the same
        visualizations as in \pr{fig:insrotorsquare}, viewed from the
        negative $y$-axis.  }
\end{figure}

The top and side views of simulation snapshots are shown in
\pr{fig:insrotorsquare} and \ref{fig:insrotorsquareside}, where
subfigures (A) - (D) show the magnitude of hydrodynamic traction, the
collision force on each particle, and the collision force
$\gamma_\ell$ on each constraint, at times $t=54\eta a^3/T$, $170 \eta
    a^3/T$, $300\eta a^3/T$ and $689\eta a^3/T$.

\pr{fig:insrotorsquare}(A) shows the initial stage of system
evolution, where particles driven by the torque tend to roll over each
other and generate many collisions within the sheet, but with the
activated constraints (right panel) being rather isolated instead of
forming chains.  At $t=170\eta a^3/T$, (B) shows the peak time of
collision force in this simulated process and, as shown in the right
panel, that many force chains are forming. Later, as shown in
subfigure (C), the collision force decreases because narrow regions
void of particles start to form.  The most striking feature is the
formation of strings of particles, or rollers in the vortical dynamics
parlance, along the direction of the torque $\bT$, as shown in
\pr{fig:insrotorsquare}(D).  These are reminiscent, perhaps, of the
Kelvin-Helmholtz rolls that form from flat vortex sheets and layers
\cite{Krasny1986,BS1990}.

\begin{figure}[htbp]
    \centering
    \includegraphics[width=\linewidth]{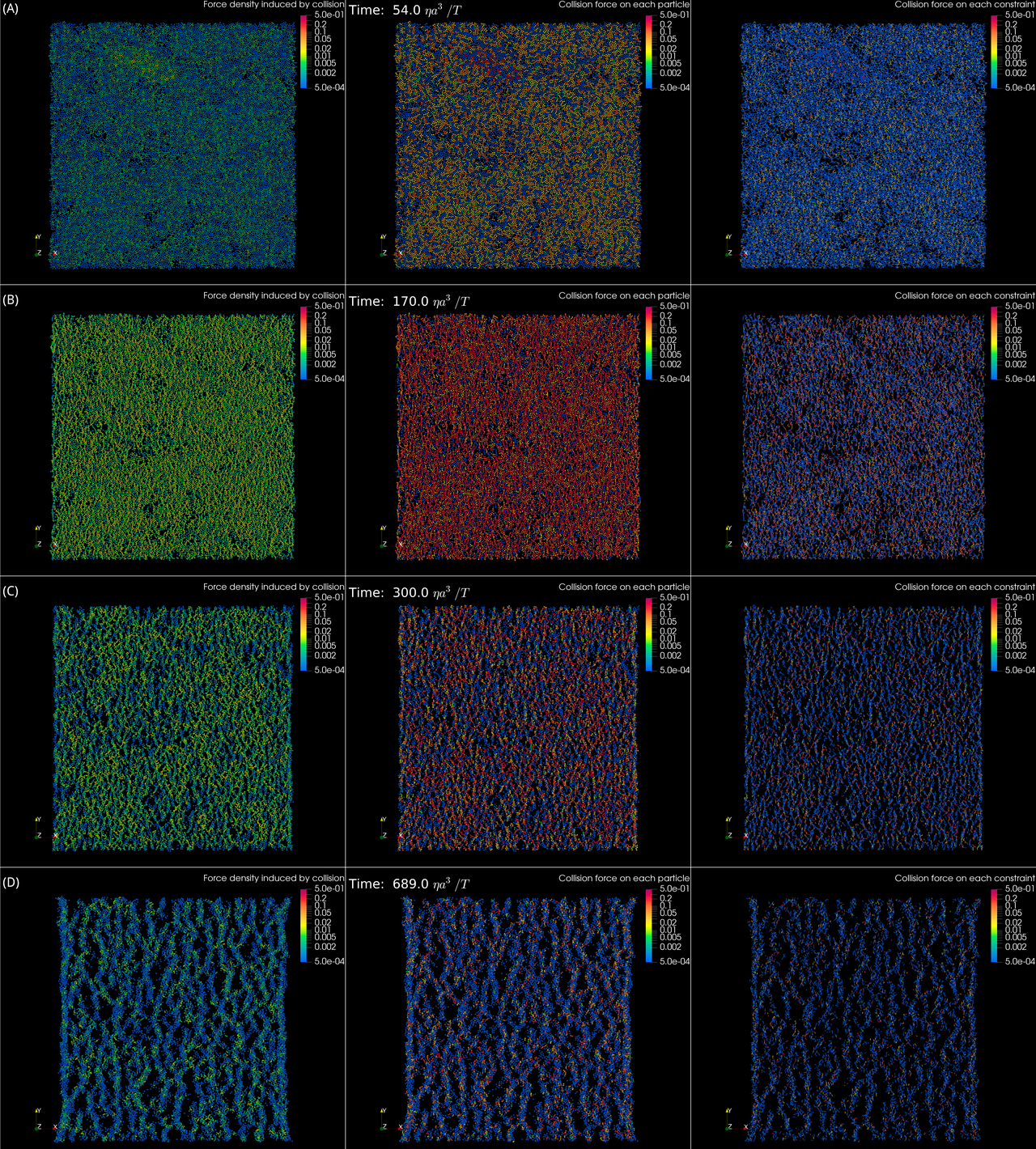}
    \caption{
        \label{fig:insrotorsquare}
        The top view of simulation snapshots for $20,000$ rotors on
        the $z=0$ plane driven by torque $\bT$ aligned with the
        $y$-axis.  (A), (B), (C), and (D) are taken at $t=54\eta
        a^3/T$, $170\eta a^3/T$, $300\eta a^3/T$ and $689\eta a^3/T$,
        respectively.  The left panels show the magnitude of
        hydrodynamic traction induced by collisions on each rotor
        surface, the middle panels show the net collision force on
        each particle, and the right panels show the collision force
        $\gamma_\ell$ on each contact constraint $\ell$.  This
        arrangement is the same as the previous example
        \pr{fig:perprotordisksnap}.  A video of this simulation is
        available in the Supplemental Material.  }
\end{figure}

The formation of these rollers can also be seen from the side view
\pr{fig:insrotorsquaresideclip}, where the arrangement is similar to
\pr{fig:insrotorsquareside} but only 10\% of particles close to the
edge are shown.  Inside each roller, the hydrodynamic traction
distributed across the rotor surface induced by the torque $\bT$ and
collisions can be clearly seen.  The formation of these chains of
rotors are related to the flow generated by the torques.  Since all
rotors are rotating along the same $+y$ direction, a global flow is
induced towards the $+x$ direction above the sheet, and towards the
$-x$ direction below the sheet.  This causes a jump of fluid velocity
across the sheet.  Further, because collectively the sheet maintains
the thin layer geometry as shown in the side view
\pr{fig:insrotorsquareside}, this jump of fluid velocity persists for
a long time, and keeps driving the formation of chains. Again, this is
similar to the Kelvin-Helmholtz instability, where the fluid velocity
jumps across a vortical layer separating two fluids. However, for our
rotor system the Reynolds number is zero, rather than infinity, and
reflects the precise balance of drive and dissipation, rather than of
dissipationless conservation laws and Hamiltonian structure (though
see \cite{lushiPeriodicChaoticOrbits2015}). A complete investigation
of these analogies is beyond the scope of this work on numerical
methods , and we leave it for a future study.

\begin{figure}[htbp]
    \centering
    \includegraphics[width=\linewidth]{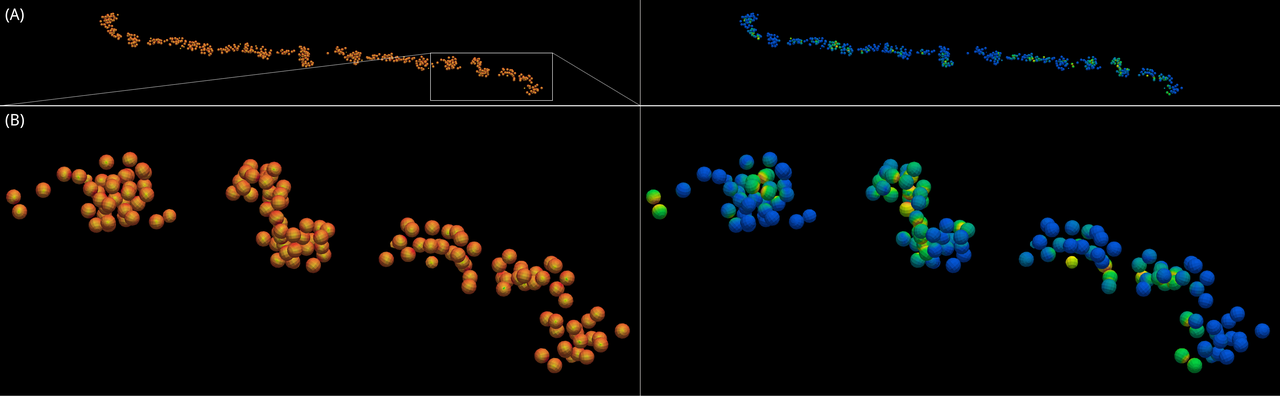}
    \caption{
        \label{fig:insrotorsquaresideclip}
        The side view of simulation snapshots at $t=600\eta a^3/T$,
        showing only those particles close to the bottom edge of the
        structure shown in \pr{fig:insrotorsquare}.  The left panels in
        (A) and (B) show the hydrodynamic traction induced by the torque
        $\bT$ on each particle, and the right panels show the hydrodynamic
        traction induced by collisions.  The colormap is the same as the
        left panels in \pr{fig:insrotorsquare}.  }
\end{figure}

\section{Conclusions}
%\label{sec:conclusions}
% Summary of collision algorithm
In this work we described a computational framework for simulating
particulate Stokes suspensions.  A key component is the collision
resolution algorithm we extended from the LCP method for underdamped
(inertial) granular flow \cite{anitescu1996,anitescu1999} to
overdamped Stokes suspensions.  The LCP is constructed at every
time-step based on the non-overlapping geometric constraints and
coupled to an explicit time-stepping scheme.  The LCP is then
converted to a CQP, utilizing the fact that the mobility matrix
$\bMcal$ is SPD, which is efficiently solved with the BBPGD method.
This collision resolution algorithm addresses two important drawbacks
in traditional collision resolution methods based on pairwise
repulsive potentials: (i) the temporal stiffness induced by repulsive
potentials, and (ii) the particles becoming effectively soft since the
repulsive potentials cannot be infinitely stiff.  This collision
resolution method does not require explicit construction of the
mobility matrix $\bMcal$.  Any mobility solver that is able to compute
$\bUcal$ with given force $\bFcal$ can be used within this collision
resolution algorithm as long as $\bMcal$ is kept SPD.  Further, the
particles do not have to be spherical \cite{Yan_spherocylinder_2019}.

We then demonstrated the application of this method to suspensions of
spherical particles, where the mobility solver is based on a new
second-kind BI equation
\cite{Corona_Greengard_Rachh_Veerapaneni_2017}.  In particular, VSH
expansions \cite{Corona_Veerapaneni_2018} are utilized to maintain
high accuracy of the BI operators for close pairs of spheres.
Consequently, this specialized mobility solver is well-conditioned in
all test cases, even when particles are very close.  The stability and
scalability of our algorithm is demonstrated in \pr{sec:results},
where we implemented these methods with full \texttt{MPI} and
\texttt{OpenMP} parallelism.  In the sedimentation and rotation tests,
time-stepping remains stable although very large time-steps are
purposefully used to demonstrate stability, even when particles move
about $10\sim20\%$ of their radii within one time-step.  In scalability
benchmarks, systems of up to $8\times 10^4$ spheres on 1792 cores are
demonstrated.  We believe the code can be successfully scaled to much
larger systems on larger machines.  Finally, in \pr{sec:application},
we demonstrated the application of our method to suspensions of driven
rotors, which illustrate the collectively-induced system-scale
rotational motions, intricate microscopic collision networks, and a
Kelvin-Helmholtz-like instability.

In comparison to a few other methods based on geometric constraints
\cite{Maury_2006,lu_contact-aware_2017}, a key advantage of our method
is that the collision force between each collision pair is
individually computed and recorded.  This preserves the entire
collision force network information, which is necessary for computing
the system collision stress.  Our approach requires computing the
collision force as a separate mobility problem, instead of embedding
the minimization problem into the mobility solver, as done by
\citet{lu_contact-aware_2017,lu_2018parallel} in their work on
deformable bodies (note that while the LCP collision resolution method
is derived for rigid particles, it does allow the particles to deform
outside this collision resolution stage).  Therefore, our method has a
higher cost because the dense mobility matrix $\bMcal$ appears in the
matrix $\bA$ of the \pr{lcp:LCPdef}. This extra cost can potentially
be reduced by the matrix-splitting subspace optimization method
\cite{robinson_subspace_2013}, where in most stages of the
minimization process only a block-diagonal part of $\bMcal$ is used.
We leave this to future work.

In other work we have demonstrated the applicability of our collision
resolution scheme to evolving assemblies of Brownian spherocylinders
\cite{Yan_spherocylinder_2019}.  
There, we simulated dynamics of a system of growing and dividing cells, 
where cell sizes increase when they grow and decrease when they
divide, and steric interactions are central to how the ``colony''
grows.  Similarly, this computational framework can be directly
applied to many other interesting physics and engineering problems,
such as confined suspensions of swimmers
\cite{reighSwimmingCageLowReynoldsnumber2017} and cell packing in
biofilms \cite{hartmannEmergenceThreedimensionalOrder2018}.

\section{Acknowledgements}
We thank E. Lushi for useful conversations.
EC and SV acknowledge support from NSF under grants DMS-1454010 and DMS-1719834.
DM acknowledges support from the Office of Naval Research under award number N00014-17-1-2451 and Simons Foundation/SFARI(560651, AB).
The work of SV was also supported by the Flatiron Institute, a division of the Simons Foundation.
MJS acknowledges the support of NSF Grants {DMR-1420073} (NYU MRSEC), {DMS-1463962}, and {DMS-1620331}.

Our implementation of this framework will be released on GitHub (https://github.com/wenyan4work/SphereSimulator) as an open-source software following the publication of this article.

\appendix
\section{The VSH expansion of traction operator}
\label{appsec:traction}

The first row of \pr{eq:tracgnm}:
\begin{align}
    g_{nm}^{rr}      & =\left(n {f_{nm}^W}'(r)-(n+1) {f_{nm}^V}'(r)\right) Y_n^m({\theta},{\phi})                                                                                                                      \\
    g_{nm}^{r\theta} & = \frac{Y_n^m({\theta},{\phi}) (-m (n+2) {f_{nm}^V}(r) \cot ({\theta})+m (n-1) {f_{nm}^W}(r) \cot ({\theta})+i m {f_{nm}^X}(r) \csc ({\theta}))}{r}\nonumber                                    \\
                     & \quad-\frac{e^{-i {\phi}} \sqrt{-m (m+1)+n^2+n} ((n+2) {f_{nm}^V}(r)-n {f_{nm}^W}(r)+{f_{nm}^W}(r)) Y_n^{m+1}({\theta},{\phi})}{r}                                                              \\
    g_{nm}^{r\phi}   & = \frac{m e^{-i {\phi}} \csc ({\theta}) (\sin ({\phi})-i \cos ({\phi})) Y_n^m({\theta},{\phi}) ((n+2) {f_{nm}^V}(r)-n {f_{nm}^W}(r)+{f_{nm}^W}(r)-i {f_{nm}^X}(r) \cos ({\theta}))}{r}\nonumber \\
                     & \quad-\frac{e^{-i {\phi}} {f_{nm}^X}(r) \sqrt{-m (m+1)+n^2+n} Y_n^{m+1}({\theta},{\phi})}{r}
\end{align}
The second row:
\begin{align}
    g_{nm}^{\theta r}     & =m \csc ({\theta}) Y_n^m({\theta},{\phi}) \left(\cos ({\theta}) \left({f_{nm}^V}'(r)+{f_{nm}^W}'(r)\right)-i {f_{nm}^X}'(r)\right)\nonumber                                                      \\
                          & \quad+e^{-i {\phi}} Y_n^{m+1}({\theta},{\phi}) \left(\sqrt{(n-m) (m+n+1)} {f_{nm}^V}'(r)+\sqrt{(n-m) (m+n+1)} {f_{nm}^W}'(r)\right)                                                              \\
    g_{nm}^{\theta\theta} & = \Bigg[\frac{{f_{nm}^V}(r) \left(-m \left(\csc ^2({\theta})-m \cot ^2({\theta})\right)-n-1\right)}{r}+\frac{{f_{nm}^W}(r) \left(m^2 \cot ^2({\theta})-m \csc ^2({\theta})+n\right)}{r}\nonumber \\
                          & \quad -\frac{i (m-1) m {f_{nm}^X}(r) \cot ({\theta}) \csc ({\theta})}{r}\Bigg]Y_n^m({\theta},{\phi})\nonumber                                                                                    \\
                          & \quad +\left(\frac{e^{-i {\phi}} \sqrt{-m (m+1)+n^2+n} \csc ({\theta}) ((2 m+1) \cos ({\theta}) ({f_{nm}^V}(r)+{f_{nm}^W}(r))-i m {f_{nm}^X}(r))}{r}\right)Y_n^{m+1}({\theta},{\phi})\nonumber   \\
                          & \quad +\left(\frac{e^{-2 i {\phi}} \sqrt{(m-n) (m-n+1) (m+n+1) (m+n+2)} ({f_{nm}^V}(r)+{f_{nm}^W}(r))}{r}\right)Y_n^{m+2}({\theta},{\phi})                                                       \\
    g_{nm}^{\theta\phi}   & = \frac{m Y_n^m({\theta},{\phi}) ({f_{nm}^X}(r)+(m-1) \csc ({\theta}) ({f_{nm}^X}(r) \csc ({\theta})+i \cot ({\theta}) ({f_{nm}^V}(r)+{f_{nm}^W}(r))))}{r} \nonumber                             \\
                          & \quad +\Bigg[\frac{i m e^{-i {\phi}} {f_{nm}^V}(r) \sqrt{-m (m+1)+n^2+n} \csc ({\theta})}{r}+\frac{i m e^{-i {\phi}} {f_{nm}^W}(r) \sqrt{-m (m+1)+n^2+n} \csc ({\theta})}{r}\nonumber            \\
                          & \quad-\frac{e^{-i {\phi}} {f_{nm}^X}(r) \sqrt{-m (m+1)+n^2+n} \cot ({\theta})}{r}\Bigg]Y_n^{m+1}({\theta},{\phi})
\end{align}
The third row:
\begin{align}
    g_{nm}^{\phi r}     & =m \csc ({\theta}) Y_n^m({\theta},{\phi}) \left(\cos ({\theta}) {f_{nm}^X}'(r)+i \left({f_{nm}^V}'(r)+{f_{nm}^W}'(r)\right)\right)\nonumber                                          \\
                        & \quad+e^{-i {\phi}} \sqrt{(n-m) (m+n+1)} {f_{nm}^X}'(r) Y_n^{m+1}({\theta},{\phi})                                                                                                   \\
    g_{nm}^{\phi\theta} & =\Bigg[\frac{i (m-1) m {f_{nm}^V}(r) \cot ({\theta}) \csc ({\theta})}{r}+\frac{i (m-1) m {f_{nm}^W}(r) \cot ({\theta}) \csc ({\theta})}{r}\nonumber                                  \\
                        & \qquad+\frac{m {f_{nm}^X}(r) \csc ^2({\theta}) (m \cos (2 {\theta})+m-2)}{2 r}\Bigg] Y_n^m(\theta,\phi)\nonumber                                                                     \\
                        & \quad\Bigg[\frac{i m e^{-i {\phi}} {f_{nm}^V}(r) \sqrt{-m (m+1)+n^2+n} \csc ({\theta})}{r}+\frac{i m e^{-i {\phi}} {f_{nm}^W}(r) \sqrt{-m (m+1)+n^2+n} \csc ({\theta})}{r}\nonumber  \\
                        & \qquad+\frac{(2 m+1) e^{-i {\phi}} {f_{nm}^X}(r) \sqrt{-m (m+1)+n^2+n} \cot ({\theta})}{r}\Bigg] Y_n^{m+1}(\theta,\phi)\nonumber                                                     \\
                        & \quad\Bigg[ \frac{i m e^{-i {\phi}} {f_{nm}^V}(r) \sqrt{-m (m+1)+n^2+n} \csc ({\theta})}{r}+\frac{i m e^{-i {\phi}} {f_{nm}^W}(r) \sqrt{-m (m+1)+n^2+n} \csc ({\theta})}{r}\nonumber \\
                        & \qquad+\frac{(2 m+1) e^{-i {\phi}} {f_{nm}^X}(r) \sqrt{-m (m+1)+n^2+n} \cot ({\theta})}{r} \Bigg]Y_n^{m+2}(\theta,\phi)                                                              \\
    g_{nm}^{\phi\phi}   & =\Bigg[\frac{{f_{nm}^V}(r) \left(-(m-1) m \csc ^2({\theta})-m-n-1\right)}{r}+\frac{{f_{nm}^W}(r) \left(m \left(\cot ^2({\theta})-m \csc ^2({\theta})\right)+n\right)}{r}\nonumber    \\
                        & \qquad+\frac{i (m-1) m {f_{nm}^X}(r) \cot ({\theta}) \csc ({\theta})}{r} \Bigg] Y_n^m(\theta,\phi)\nonumber                                                                          \\
                        & \quad\Bigg[ \frac{e^{-i {\phi}} {f_{nm}^V}(r) \sqrt{-m (m+1)+n^2+n} \cot ({\theta})}{r}+\frac{e^{-i {\phi}} {f_{nm}^W}(r) \sqrt{-m (m+1)+n^2+n} \cot ({\theta})}{r}\nonumber         \\
                        & \qquad +\frac{i m e^{-i {\phi}} {f_{nm}^X}(r) \sqrt{-m (m+1)+n^2+n} \csc ({\theta})}{r} \Bigg] Y_n^{m+1}(\theta,\phi)\nonumber                                                       \\
\end{align}

\section{The singularity of velocity close to the disk edge of Stokes rotlets}
\label{appsec:rotletdisk}
\subsection{Geometry and setup}
We consider a domain $D$ of a disk on $z=0$ plane:
\begin{align}
    D = \{(x,y)|x^2+y^2<R^2\}
\end{align}
We denote the number density of particles within this disk as $n$.  A
constant torque $\bT$ toward $+z$ direction is exerted on each
particle.  We assume that each particle follows the fluid velocity
induced by other particles.  Each particle induces a rotational fluid
flow $u_j$, as given by the Stokes rotlet velocity field:
\begin{align}
    u_j = \frac{\epsilon_{jlm}}{8\pi\mu}\frac{r_m}{r^3}T_l,
\end{align}
where $\epsilon_{jlm}$ is the Levi-Civita tensor.  $\br$ is the vector
pointing from the rotlet to the point where $u_j$ is evaluated.

Due to the symmetry of this disk $D$, we consider a point $(s,0)$ on
the $x$-axis.  The velocity $\bu$ at this point is aligned with the
$y$-axis, given by an integral over the rotlets on this entire disk:
\begin{align}
    \frac{8\pi\mu }{n T} u(s)= \fint_0^R 2\pi r \int_0^{2\pi}
    \frac{s-r\cos\theta}{\left((s-r\cos\theta)^2+r^2\sin^2\theta\right)^3}
    d\theta dr,
\end{align}
where $r,\theta$ is the cylindrical coordinate system used to denote the rotlet point in this disk of radius $R$.
$\fint$ denotes the principal value of this integral, because this integral involves a high order of singularity at the point $(r,0)$.

The integral over $\theta$ can be computed to get the closed form:
\begin{align}
    F(r) = 2 \left( \frac{K\left(\frac{4 r s}{(r+s)^2}\right)}{s (r+s)}+\frac{E\left(\frac{4 r s}{(r+s)^2}\right)}{s (s-r)} \right),
\end{align}
where $K(x)$ is the complete elliptic integral of the first kind \cite{EIK}, and $E(x)$ is the complete elliptic integral of the second kind \cite{EIE}.
$K(x)$ is singular at $x=1$, i.e., $r=s$.
So the two terms of $F(r)$ are both singular at $r=s$.
As a result, the velocity $u(s)$ becomes:
\begin{align}
    \frac{8\pi\mu }{n T} u(s)= \fint_0^R F(r) 2\pi r dr	,
\end{align}
where $0<s<R$.
The principal value can be computed as this limit as $\delta\to0^+$:
\begin{align}
    \frac{8\pi\mu }{n T} u(s) = \lim_{\delta\to 0+}
    \left(\int_0^{s-\delta} F(r) 2\pi r dr + \int_{s+\delta}^R F(r)
    2\pi r dr \right).
\end{align}

\subsection{Principal value and singularity}
Now we discuss the singularity in a small region $[s-a,s+b]$ around $s$, where $s-a<s-\delta<s<s+\delta<s+b$.
In this small region, we use the asymptotic expansion of the integrand $r F(r)$ around $r=s$ \cite{DLMF}:
\begin{align}
    rF(r) & \approx \frac{-2 \log (s-r)-\log \left(\frac{1}{4 s^2}\right)-4+4 \log (2)}{2 s}-\frac{2}{r-s} \quad\text{ when } r<s \\
          & \approx \frac{-2 \log (r-s)-\log \left(\frac{1}{4 s^2}\right)-4+4 \log (2)}{2 s}-\frac{2}{r-s} \quad\text{ when } r>s
\end{align}

In the small region $[s-a,s+b]$, the integral can be asymptotically computed:
\begin{align}
    \fint_{s-a}^{s+b} rF(r)dr & = \int_{s-a}^{s-\delta} F(r) r dr + \int_{s+\delta}^{s+b} F(r) r dr                                                                                                                         \\
                              & \approx \frac{(\delta-a) \log \left(\frac{1}{s^2}\right)+4 s \log \left(\frac{a}{\delta}\right)+(\log (64)-2) (a-\delta)-2 a \log (a)+2 \delta \log (\delta)}{2 s}\nonumber                 \\
                              & + \frac{(b-\delta) \left(-\log \left(\frac{1}{s^2}\right)-2+\log (64)\right)+4 s \log \left(\frac{\delta}{b}\right)-2 b \log (b)+2 \delta \log (\delta)}{2 s}                               \\
                              & =\frac{(\log (64)-2) (a+b-2 \delta)-(a+b) \log \left(\frac{1}{s^2}\right)+4 s \log \left(\frac{a}{b}\right)-2 a \log (a)-2 b \log (b)+2 \delta \log \left(\frac{\delta^2}{s^2}\right)}{2 s}
\end{align}
The limit as $\delta\to 0$ exists:
\begin{align}
    \lim_{\delta\to 0} & \fint_{s-a}^{s+b} rF(r)dr = 2 \log \left(\frac{a}{b}\right)\nonumber                                                                                                               \\
                       & + \frac{-\frac{1}{2} a \log \left(\frac{1}{s^2}\right)-a+\frac{1}{2} a \log (64)-a \log (a)-\frac{1}{2} b \log \left(\frac{1}{s^2}\right)-b+\frac{1}{2} b \log (64)-b \log (b)}{s}
\end{align}
This gives the behavior of the singularity around the point $(s,0)$.

When the point $(s,0)$ is close to the edge of the disk, i.e., $s\to R^-$:
\begin{align}
    b=R-s \to 0^+
\end{align}
The behavior of this pole can be calculated as the limit of $b\to 0^+$:
\begin{align}
\left(\lim_{\delta\to 0} \int_{s-a}^{s+b} rF(r)dr\right) = -2\log b -\frac{1}{R} b\log b + O(b) +C(a) \quad \text{ as }b\to 0^+ 
\end{align}
where we have used $s\to R^-$.
$O(b)$ is of order $b$ or higher.
$C(a)$ is a function involving $a$ and $R$ only.

In sum, for a target point at $(s,0)$ close to the disk edge, the velocity shows a logarithm singularity dominated by the $\log b$ term:
\begin{align}
    u = A \log \left(R-s\right) + B + O\left(\frac{R-s}{R}\log(R-s)\right),
\end{align}
where $A$ and $B$ are two constants to be determined.

\bibliographystyle{elsarticle-num-names}
\section*{\refname}
\bibliography{ref.bib}

\end{document}